\documentclass[12pt,a4paper]{article}
\usepackage{amssymb}
\usepackage{amsfonts}
\usepackage{amsmath}
\usepackage{float}
\usepackage{latexsym}
\usepackage{cite}
\begin{document}

\title{{\bf Perturbations against a Q-ball: Charge,\\ energy, and additivity property}}

\author{
Mikhail~N.~Smolyakov$^{1,2}$
\\
$^1${\small{\em Skobeltsyn Institute of Nuclear Physics, Lomonosov Moscow
State University,
}}\\
{\small{\em Moscow 119991, Russia}}\\
$^2${\small{\em
Institute for Nuclear Research of the Russian Academy
of Sciences,}}\\
{\small{\em 60th October Anniversary prospect 7a, Moscow 117312,
Russia
}}}

\date{}
\maketitle

\begin{abstract}
In the present paper, perturbations against a Q-ball solution are considered. It is shown that if we calculate the U(1) charge and the energy of the modes, which are solutions to linearized equations of motion, up to the second order in perturbations, we will get incorrect results. In particular, for the time-dependent modes we will obtain nonzero terms, which explicitly depend on time, indicating the nonconservation over time of the charge and the energy. It is shown that, as expected, this problem can be resolved by considering nonlinear equations of motion for the perturbations, providing second-order corrections to the solutions of linearized equations of motion. It turns out that contributions of these corrections to the charge and the energy can be taken into account without solving explicitly the nonlinear equations of motion for the perturbations. It is also shown that the use of such nonlinear equations not only recovers the conservation over time of the charge and the energy but also results in the additivity of the charge and the energy of different modes forming the perturbation.
\end{abstract}

\section{Introduction}
It is well known that in classical field theory charge and energy are conserved over time if the corresponding equations of motion are satisfied. Of course, the same should be valid if we consider, for example, a stationary background solution and a small perturbation against it: charge and energy of the whole system are conserved over time, so we expect that the correctly defined charge and energy of the perturbation are also conserved over time. As a simple and well-known example, one can recall perturbations against the kink solution, satisfying the corresponding linearized equation of motion (see, for example, \cite{Rajaraman:1982is}). The energy of the modes, which is quadratic in perturbations, does not depend on time, ensuring its conservation over time.

Usually, perturbations against some background solution are supposed to satisfy the linearized equations of motion, whereas the charge and the energy of the perturbations are calculated up to the terms quadratic in perturbations. Meanwhile, since the conservations laws
\begin{equation}
\frac{dQ}{dt}=0,\qquad \frac{dE}{dt}=0
\end{equation}
are valid whenever the equations of motion are satisfied, if the equations of motion for perturbations are linear, then in the general case one can expect that the charge and the energy are conserved only in the linear order in perturbations, but not in the quadratic order in perturbations. Below it will be shown explicitly that exactly this situation is realized if we consider a Q-ball as a background solution.

A solution to this problem is obvious and simple: in order to have the conservation over time of the charge and the energy up to the second order in perturbations, in the general case one has to consider equations of motion for the perturbations also up to the second order in perturbations. I will show explicitly how it works for Q-balls, revealing some interesting results in the course of the analysis. In particular, I will show that although the correct expressions for the charge and the energy of a single mode include nonlinear corrections, their contribution can be taken into account without solving explicitly the nonlinear equations of motion for perturbations. Moreover, the use of these nonlinear equations of motion not only recovers the conservation over time of the charge and the energy but also results in the additivity of the charge and the energy of different modes forming the perturbation. Exactly the same effect has recently been observed in the case of the nonlinear Schr\"{o}dinger equation \cite{Smolyakov}.

The paper is organized as follows. In Section~2 the basic setup and notations are introduced. In Section~3 perturbations against a Q-ball in the linear approximation are considered. In Section~4 it is shown explicitly that the use of the linear approximation for perturbations is not sufficient to provide correct expressions for the charge and the energy of the modes up to the second order in perturbations. In Section~5 perturbations in the nonlinear approximation (including the corrections quadratic in perturbations) are discussed and correct expressions for the charge and the energy of the modes are obtained. It is also shown that the charge and the energy of a single mode can be calculated without solving explicitly the nonlinear equations of motion for perturbations. In Section~6 it is shown that the additivity property is valid for the charge and the energy of oscillation modes. In Section~7 a simple explicit formula, which allows one to isolate a single oscillation mode from the linear part of the perturbation, is presented. In Section~8 an explicit example of the model with logarithmic scalar field potential is discussed. In the Conclusion the obtained results are briefly discussed. The auxiliary materials are collected in four appendices.

\section{Setup}
Let us start with the standard action, describing a complex scalar field $\phi$ in the flat $(d+1)$-dimensional space-time with $d\ge 1$, in the form
\begin{equation}\label{action}
S=\int\left(\dot\phi^*\dot\phi-\sum_{l=1}^{d}\partial_{l}\phi^*\partial_{l}\phi-V(\phi^*\phi)\right)dtd^{d}x,
\end{equation}
where $\dot\phi=\partial_{t}\phi$. The standard ansatz for a Q-ball has the form \cite{Rosen0,Coleman:1985ki}
\begin{equation}\label{qballsolution}
\phi_{0}(t,\vec x)=e^{i\omega t}f(r).
\end{equation}
Here $r=\sqrt{{\vec x}^{2}}$ and $f(r)$ is a real function that is supposed to have no nodes (without loss of generality, we can set $f(r)>0$ for any $r$) and to satisfy the boundary conditions
\begin{equation}
\partial_{r}f(r)|_{r=0}=0,\qquad \lim\limits_{r\to\infty}f(r)=0.
\end{equation}
In this case, the function $f(r)$ satisfies the equation
\begin{equation}\label{eqqball}
\omega^{2}f+\Delta f-\frac{dV}{d(\phi^{*}\phi)}\biggl|_{\phi^{*}\phi=f^{2}}f=0,
\end{equation}
where $\Delta=\sum\limits_{l=1}^{d}\partial_{l}^{2}$. Of course, the function $f(r)$ depends on $\omega$ (more precisely, on $\omega^{2}$).

The U(1) charge, corresponding to the theory with action \eqref{action}, is defined in the standard way as
\begin{equation}
Q=i\int\left(\phi\dot\phi^*-\phi^*\dot\phi\right)d^{d}x.
\end{equation}
In particular, for the Q-ball we get
\begin{equation}\label{qballcharge}
Q_{0}=2\omega\int f^{2}d^{d}x.
\end{equation}
The energy of this system has the form
\begin{equation}
E=\int\left(\dot\phi^*\dot\phi+\sum_{l=1}^{d}\partial_{l}\phi^*\partial_{l}\phi+V(\phi^*\phi)\right)d^{d}x.
\end{equation}

\section{Perturbations in the linear approximation}\label{pertsection}
Now let us consider small perturbations against background solution \eqref{qballsolution}, of the form
\begin{equation}
\phi(t,\vec x)=e^{i\omega t}\bigl(f(r)+\varphi(t,\vec x)\bigr).
\end{equation}
In the linear approximation, the perturbation $\varphi(t,\vec x)$ satisfies the equation of motion
\begin{align}\nonumber
e^{i\omega t}\left(\omega^{2}\varphi-2i\omega\dot\varphi-\ddot\varphi+\Delta\varphi-\frac{dV}{d(\phi^{*}\phi)}\biggl|_{\phi^{*}\phi=f^{2}(r)}\varphi\right)\\
-\frac{d^{2}V}{d(\phi^{*}\phi)^{2}}\biggl|_{\phi^{*}\phi=f^{2}(r)}\left(\phi_{0}^{*}\phi_{0}^{}e^{i\omega t}\varphi+\phi_{0}^{2}e^{-i\omega t}\varphi^{*}\right)=0,
\end{align}
where $\phi_{0}$ is given by \eqref{qballsolution}, which can be rewritten as
\begin{equation}\label{lineqgeneral}
\omega^{2}\varphi-2i\omega\dot\varphi-\ddot\varphi+\Delta\varphi-U\varphi-S(\varphi+\varphi^{*})=0,
\end{equation}
where
\begin{equation}\label{UGdef}
U(r)=\frac{dV}{d(\phi^{*}\phi)}\biggl|_{\phi^{*}\phi=f^{2}(r)},\qquad S(r)=\frac{d^{2}V}{d(\phi^{*}\phi)^{2}}\biggl|_{\phi^{*}\phi=f^{2}(r)}f^{2}(r).
\end{equation}
The standard ansatz for perturbations, which passes through the linearized equation of motion, takes the form \cite{Anderson:1970et,MarcVent}
\begin{equation}\label{substgeneral}
\varphi(t,\vec x)=\alpha\left(a(\vec x)e^{i\rho t}+b(\vec x)e^{-i\rho^{*}t}\right),
\end{equation}
where $a(\vec x)$ and $b(\vec x)$ are complex functions, $\rho$ is a complex parameter, and $\alpha\ll 1$ is a real parameter. Of course, this ansatz does not describe all types of perturbations, even if we do not take into account possible exotic cases such as the mode of the form
\begin{equation}
\varphi(t,\vec x)\sim itf+\frac{df}{d\omega},
\end{equation}
which corresponds to the change of the Q-ball frequency $\omega$. As an example, one can also recall the mode of the form
\begin{equation}\label{Lorentz0}
\varphi(t,\vec x)\sim t\partial_{j}f+i\omega x^{j}f,
\end{equation}
which corresponds to the Lorentz symmetry of the theory. However, ansatz \eqref{substgeneral} covers the most important types of perturbations, whereas the mode defined by \eqref{Lorentz0} will be considered separately.

For ansatz \eqref{substgeneral}, the equations of motion take the form
\begin{align}\label{lineq1intro}
\hat L_{1}\xi-2\omega\rho\eta-\rho^{2}\xi=0,\\ \label{lineq2intro}
\hat L_{2}\eta-2\omega\rho\xi-\rho^{2}\eta=0,
\end{align}
where
\begin{equation}
\xi(\vec x)=a(\vec x)+b^{*}(\vec x),\qquad \eta(\vec x)=a(\vec x)-b^{*}(\vec x)
\end{equation}
and
\begin{align}\label{L1eq}
&\hat L_{1}=-\Delta+U(r)+2S(r)-\omega^{2},\\ \label{L2eq}
&\hat L_{2}=-\Delta+U(r)-\omega^{2}.
\end{align}

Using linearized equations of motion \eqref{lineq1intro} and \eqref{lineq2intro}, it is possible to show that the following relation fulfills
\begin{equation}
\rho^{2}={\rho^{*}}^2;
\end{equation}
see the detailed proof in \cite{Panin:2016ooo}. This equation has two obvious solutions: $\rho=\gamma$ and $\rho=-i\gamma$, where $\gamma$ is a real constant. The first solution, $\rho=\gamma$, stands for the oscillation modes, which have the form
\begin{equation}
\varphi(t,\vec x)=\alpha\left(a(\vec x)e^{i\gamma t}+b(\vec x)e^{-i\gamma t}\right).
\end{equation}
Without loss of generality, we can set $\gamma>0$ in this case. Note that perturbations of a simpler form, say, of the form
\begin{equation}
\varphi(t,\vec x)=\alpha a(\vec x)e^{i\gamma t},
\end{equation}
do not pass through linearized equation of motion \eqref{lineqgeneral}. The second solution, $\rho=-i\gamma$, leads to
\begin{equation}\label{instmode}
\varphi(t,\vec x)=\alpha(a(\vec x)+b(\vec x))e^{\gamma t}=\alpha(u(\vec x)+iv(\vec x))e^{\gamma t},
\end{equation}
where $u(\vec x)$ and $v(\vec x)$ are real functions. Of course, such modes exist only for classically unstable Q-balls and describe the initial stage of their classical decay.\footnote{The criterion of classical stability for Q-balls was established in \cite{Friedberg:1976me}; see also \cite{Kumar:1979sq,GSS,Grillakis,LeePang}. The proof along the lines of the Vakhitov-Kolokolov method \cite{VK,Kolokolov}, based on the use of only the linearized equations of motion for perturbations, can be found in \cite{Panin:2016ooo}.} Usually, a classically unstable Q-ball has only one such mode in the spectrum of perturbations (this statement is not proved; however, I am not aware of any exceptions), and this mode is spherically symmetric (this topic will be discussed later). Technically, time-independent modes (such as translational modes) can be considered as instability modes with $\gamma=0$ , but for illustration purposes below I will consider these modes separately.

From \eqref{instmode} it follows that formally the case $\rho=-i\gamma$ should be considered separately, because we cannot isolate the terms $\sim e^{i\rho t}$ and $\sim e^{-i\rho^{*}t}$ in the linearized equation of motion. However, the correct equations for $u$ and $v$ can be obtained directly from Eqs.~\eqref{lineq1intro} and \eqref{lineq2intro} if we suppose that $\xi$ is purely real and $\eta$ is purely imaginary and set $u=\xi$, $v=-i\eta$.

It is interesting to mention that when a Q-ball passes (by changing its frequency $\omega$) from the stable region $\frac{dQ_{0}}{d\omega}<0$ to the unstable region $\frac{dQ_{0}}{d\omega}>0$,\footnote{Usually, there exists the instability mode for Q-balls with $\frac{dQ_{0}}{d\omega}>0$, although the classical stability criterion only states that there are no instability modes if $\frac{dQ_{0}}{d\omega}<0$ and if there exists only one negative eigenvalue of the operator $\hat L_{1}$.} the oscillation mode turns into the instability mode (the explicit manifestation of this effect will be presented in Section~8).\footnote{In $(1+1)$-dimensional case, this effect was observed numerically in \cite{Belova}.} To demonstrate it, let us consider a Q-ball in the vicinity of $\frac{dQ_{0}}{d\omega}=0$, for which $\frac{|\rho|}{\omega}\ll 1$ is supposed to fulfill, and represent the perturbation as
\begin{align}\label{pertcomplex1}
&\xi=c\left(\rho\frac{df}{d\omega}+\rho^{3}c_{1}(\vec x)\right),\\ \label{pertcomplex2}
&\eta=c\left(f+\rho^{2}c_{2}(\vec x)\right),
\end{align}
where $c$ is a complex constant and $c_{1}(\vec x)$ and $c_{2}(\vec x)$ are complex functions. The form of perturbation \eqref{pertcomplex1} and \eqref{pertcomplex2} is motivated by the facts that $\hat L_{2}f=0$ (which is simply Eq.~\eqref{eqqball}) and
\begin{equation}\label{dfdomegaL1}
\hat L_{1}\frac{df}{d\omega}=2\omega f;
\end{equation}
the latter relation can be obtained by differentiating Eq.~\eqref{eqqball} with respect to $\omega$. Using ansatz \eqref{pertcomplex1} and \eqref{pertcomplex2} and linearized equations of motion \eqref{lineq1intro} and \eqref{lineq2intro}, for $|\rho|\to 0$ we can get the relation (see the derivation in Appendix~A)
\begin{equation}\label{stinstsmallrho}
\frac{1}{2}\frac{dQ_{0}}{d\omega}+\rho^{2}\int\left[\left(\frac{df}{d\omega}\right)^{2}+c_{2}^{*}\hat L_{2}c_{2}\right]d^{d}x\simeq 0.
\end{equation}
Now recall that $\hat L_{2}f=0$, i.e., the function $f$ is the eigenfunction of the operator $\hat L_{2}$ with the zero eigenvalue. Since $f(r)>0$ for any $r$, it is the eigenfunction of the lowest eigenstate of the operator $\hat L_{2}$ (for example, for $d=3$ it corresponds to the $1s$ level in the spherically symmetric quantum mechanical potential $U(r)-\omega^{2}$), so all the other eigenvalues of $\hat L_{2}$ are larger than zero, leading to
\begin{equation}\label{L2positive}
\int c_{2}^{*}\hat L_{2}c_{2}\,d^{d}x\ge 0.
\end{equation}
Thus, for $\frac{dQ_{0}}{d\omega}<0$ we have $\rho^{2}>0$, leading to $\rho=\gamma$ and
\begin{equation}
\varphi\simeq\alpha\frac{c}{2}\left(\gamma\frac{df}{d\omega}+f\right)e^{i\gamma t}+\alpha\frac{c^{*}}{2}\left(\gamma\frac{df}{d\omega}-f\right)e^{-i\gamma t},
\end{equation}
whereas for $\frac{dQ_{0}}{d\omega}>0$ we have $\rho^{2}<0$, leading to $\rho=-i\gamma$ and
\begin{equation}
\varphi\simeq-i\alpha\frac{c-c^{*}}{2}\left(\gamma\frac{df}{d\omega}+if\right)e^{\gamma t}.
\end{equation}

\section{Charge and energy: examples of incorrect calculations}
Let us define the charge and the energy of the perturbation $\varphi$ up to the second order in perturbations as follows:
\begin{align}\label{chargegeneral}
&Q_{p}=Q-Q_{0}=\int\Bigl(2\omega f(\varphi+\varphi^{*})+if(\dot\varphi^{*}-\dot\varphi)+2\omega\varphi^{*}\varphi+i(\dot\varphi^{*}\varphi-\varphi^{*}\dot\varphi)\Bigr)d^{d}x,\\
\nonumber
&E_{p}=E-E_{0}=\int\Bigl(2\omega^{2}f(\varphi+\varphi^{*})+i\omega f(\dot\varphi^{*}-\dot\varphi)\\\label{energygeneral}&+(\dot\varphi^{*}-i\omega\varphi^{*})(\dot\varphi+i\omega\varphi)+\sum_{l=1}^{d}\partial_{l}\varphi^{*}\partial_{l}\varphi+
U\varphi^{*}\varphi+\frac{1}{2}S(\varphi+\varphi^{*})^{2}\Bigr)d^{d}x,
\end{align}
where $E_{0}$ is the Q-ball energy. Note that expression \eqref{chargegeneral} is exact. The calculation of charge and energy \eqref{chargegeneral} and \eqref{energygeneral} seems to be a trivial and straightforward task. Indeed, we have perturbation \eqref{substgeneral}, which is supposed to satisfy linearized equation of motion \eqref{lineqgeneral}, so it is necessary just to substitute \eqref{substgeneral} into \eqref{chargegeneral} and  \eqref{energygeneral} and to calculate the corresponding integrals. However, below we will see that, as was noted in the beginning of the paper, such an approach leads to incorrect results.

\subsection{Time-independent mode}
First, let us consider the simplest example
\begin{equation}\label{modeU1}
\varphi(t,\vec x)=i\alpha f(r),
\end{equation}
where $\alpha$ is a real dimensionless constant such that $\alpha\ll 1$, which corresponds to the global U(1) symmetry $f\to e^{i\alpha}f$ and satisfies linearized equation of motion \eqref{lineqgeneral}. Substituting \eqref{modeU1} into \eqref{chargegeneral} and \eqref{energygeneral}, we easily get for the charge
\begin{equation}
Q_{p}=2\alpha^{2}\omega\int f^{2}d^{d}x.
\end{equation}
For the energy of the perturbation, using the equation of motion for the background field $f$, we get
\begin{equation}
E_{p}=2\alpha^{2}\omega^{2}\int f^{2}d^{d}x.
\end{equation}
This result seems to be incorrect, because multiplication of the initial solution $f$ by $e^{i\alpha}$ does not change the total charge and energy, so we expect to get $Q_{p}=0$, $E_{p}=0$, but not $Q_{p}\neq 0$, $E_{p}\neq 0$.

\subsection{Modes corresponding to the Lorentz transformations}
Let us consider the mode
\begin{equation}\label{Lorentz1}
\varphi(t,\vec x)=v\left(t\partial_{j}f+i\omega x^{j}f\right),
\end{equation}
where $v\ll 1$ and $j=1,2,...,d$. This mode corresponds to the linear in $v$ term of the Taylor series of the solution
\begin{equation}
\phi(t,\vec x)=e^{i\omega\left(\frac{t+vx^{j}}{\sqrt{1-v^{2}}}\right)}f\left(x^{1},...,\frac{x^{j}+vt}{\sqrt{1-v^{2}}},...,x^{d}\right),
\end{equation}
which corresponds to a Q-ball moving with a constant speed $v$ along the $j$th direction. It is not difficult to check that \eqref{Lorentz1} satisfies linearized equation of motion \eqref{lineqgeneral}. Substituting \eqref{Lorentz1} into \eqref{chargegeneral}, we easily get
\begin{equation}
Q_{p}=v^{2}2\omega\int\left(t^{2}(\partial_{j}f)^{2}+\omega^{2}(x^{j})^{2}f^{2}+\frac{1}{2}f^{2}\right)d^{d}x,
\end{equation}
which is not conserved over time. Moreover, we expect to get $Q_{p}=0$ in this case, but not $Q_{p}\neq 0$. Analogous results can be obtained for $E_{p}$.

\subsection{Instability modes}
Now let us consider the mode
\begin{equation}\label{substdecayincorrect}
\varphi(t,\vec x)=\alpha e^{\gamma t}\Bigl(u(\vec x)+iv(\vec x)\Bigr),
\end{equation}
where $\gamma$ is real and $\gamma\neq 0$, $\alpha$ is real and $\alpha\ll 1$, and $u(\vec x)$ and $v(\vec x)$ are real functions. For mode \eqref{substdecayincorrect}, the linearized equations of motion take the form
\begin{align}\label{equv1}
\hat L_{1}u-2\omega\gamma v+\gamma^{2}u=0,\\ \label{equv2}
\hat L_{2}v+2\omega\gamma u+\gamma^{2}v=0.
\end{align}
As was noted above, the instability mode is spherically symmetric. This happens because usually the operator $\hat L_{1}$ has only one negative eigenvalue and the corresponding eigenfunction is spherically symmetric.\footnote{The existence of only one negative eigenvalue of the operator $\hat L_{1}$ is essential for the validity of the classical stability criterion $\frac{dQ_{0}}{d\omega}<0$; see, for example, \cite{Panin:2016ooo}.} To demonstrate it, let us decompose the functions $u$ and $v$ into eigenfunctions of the operator $\hat L_{1}$. In such a case, Eqs.~\eqref{equv1} and \eqref{equv2} decouple into equations for the spherically symmetric parts of $u$, $v$ and nonspherically symmetric parts of $u$, $v$ (which can be denoted as $u_{\textrm{ns}}$ and $v_{\textrm{ns}}$). Thus, from Eqs.~\eqref{equv1} and \eqref{equv2} for $u_{\textrm{ns}}$, $v_{\textrm{ns}}$ it follows that
\begin{equation}\label{instnospher1}
\gamma^{2}\int(u_{\textrm{ns}}^{2}+v_{\textrm{ns}}^{2})d^{d}x=-\int(u_{\textrm{ns}}\hat L_{1}u_{\textrm{ns}}+v_{\textrm{ns}}\hat L_{2}v_{\textrm{ns}})d^{d}x.
\end{equation}
Since the only eigenfunction of the operator $\hat L_{1}$ with negative eigenvalue is spherically symmetric, we get
\begin{equation}\label{instnospher2}
\int u_{\textrm{ns}}\hat L_{1}u_{\textrm{ns}}\, d^{d}x\ge 0.
\end{equation}
Using Eqs.~\eqref{instnospher2} and \eqref{L2positive} we get
\begin{equation}
\gamma^{2}\int(u_{\textrm{ns}}^{2}+v_{\textrm{ns}}^{2})d^{d}x\le 0,
\end{equation}
which has only a trivial solution for $\gamma^{2}>0$. This implies that at least if $\hat L_{1}$ has only one negative eigenvalue, the instability mode can be only spherically symmetric. Equation \eqref{instnospher1}, but with $u$, $v$ instead of $u_{\textrm{ns}}$, $v_{\textrm{ns}}$, also implies that in the general case $\gamma^{2}<|\lambda_{-}|$, where $\lambda_{-}$ is the negative eigenvalue of the operator $\hat L_{1}$.

Substituting \eqref{substdecayincorrect} into \eqref{chargegeneral}, we get
\begin{equation}
Q_{p}=\alpha e^{\gamma t}\int(4\omega fu+2\gamma fv)d^{d}x+\alpha^{2}e^{2\gamma t}2\omega\int(u^{2}+v^{2})d^{d}x.
\end{equation}
If we multiply Eq.~\eqref{equv2} by $f$, integrate the result over the space and use the fact that $\hat L_{2}f=0$, we will obtain
\begin{equation}\label{linvanishdecay}
\int(2\omega fu+\gamma fv)d^{d}x=0.
\end{equation}
The latter relation leads to
\begin{equation}
Q_{p}=\alpha^{2}e^{2\gamma t}2\omega\int(u^{2}+v^{2})d^{d}x.
\end{equation}
It is clear that the integral $\int d^{d}x(u^{2}+v^{2})$ is not equal to zero. Thus, the charge of the instability mode is not conserved over time. Analogously, for the energy we get\footnote{When calculating $E_{p}$, one should perform integration by parts and use Eqs.~\eqref{equv1} and \eqref{equv2} to get rid of the terms with $\sum\limits_{l=1}^{d}\partial_{l}u\partial_{l}u$, $\sum\limits_{l=1}^{d}\partial_{l}v\partial_{l}v$.}
\begin{equation}
E_{p}=\alpha^{2}e^{2\gamma t}2\omega^{2}\int(u^{2}+v^{2})d^{d}x,
\end{equation}
which also is not conserved over time.

\subsection{Oscillation modes}
And finally, let us consider the oscillation mode, which has the form
\begin{equation}
\varphi(t,\vec x)=\alpha\left(a(\vec x)e^{i\gamma t}+b(\vec x)e^{-i\gamma t}\right),
\end{equation}
where $\gamma$ is real and $\gamma\neq 0$, $\alpha$ is real and $\alpha\ll 1$, and $a(\vec x)$ and $b(\vec x)$ are complex functions. From the very beginning, it is convenient to use the notations
\begin{equation}\label{notatosc}
\xi(\vec x)=a(\vec x)+b^{*}(\vec x),\qquad \eta(\vec x)=a(\vec x)-b^{*}(\vec x).
\end{equation}
With these notations, the corresponding linearized equations of motion take the form
\begin{align}\label{lineq1}
\hat L_{1}\xi-2\omega\gamma\eta-\gamma^{2}\xi=0,\\ \label{lineq2}
\hat L_{2}\eta-2\omega\gamma\xi-\gamma^{2}\eta=0.
\end{align}
For the charge of the mode, we obtain
\begin{align}\nonumber
Q_{p}=\alpha\int\Bigl(e^{i\gamma t}(2\omega f\xi+\gamma f\eta)+e^{-i\gamma t}(2\omega f\xi^{*}+\gamma f\eta^{*})\Bigr)d^{d}x\\+
\alpha^{2}\int\left(\omega(\xi^{*}\xi+\eta^{*}\eta)+\gamma(\xi^{*}\eta+\eta^{*}\xi)+\frac{\omega}{2}\left(e^{2i\gamma t}(\xi^{2}-\eta^{2})+e^{-2i\gamma t}({\xi^{*}}^{2}-{\eta^{*}}^{2})\right)\right)d^{d}x.
\end{align}
If we multiply Eq.~\eqref{lineq2} by $f$, integrate the result over the space, and use the fact that $\hat L_{2}f=0$, we will get
\begin{equation}\label{osclinterm}
\int(2\omega f\xi+\gamma f\eta)d^{d}x=0.
\end{equation}
Thus, the terms $\sim\alpha$ vanish, and we arrive at
\begin{align}\label{chargenoncons}
Q_{p}=\alpha^{2}\int\Bigl(\omega(\xi^{*}\xi+\eta^{*}\eta)+\gamma(\xi^{*}\eta+\eta^{*}\xi)+e^{2i\gamma t}\frac{\omega}{2}(\xi^{2}-\eta^{2})+e^{-2i\gamma t}\frac{\omega}{2}({\xi^{*}}^{2}-{\eta^{*}}^{2})\Bigr)d^{d}x.
\end{align}
In the general case
\begin{equation}
\int(\xi^{2}-\eta^{2})d^{d}x=4\int ab^{*}d^{d}x\neq 0.
\end{equation}
The fact that this integral is not equal to zero is not so obvious as in the previous example of the instability mode, but it can be checked explicitly for the perturbations in the model of \cite{Rosen1}, which were examined analytically in \cite{MarcVent}. Analogous time-dependent terms can be obtained for $E_{p}$. Thus, the charge and the energy are not conserved over time.

As has been already mentioned, the origin of nonconservation of the charge and the energy, as well as of the incorrect results for the time-independent modes, is trivial. The conservation laws $\dot Q=0$ and $\dot E=0$ are valid only if the equations of motion are satisfied. Thus, if the equations of motion are satisfied only in the linear approximation, in the general case we cannot expect that the charge and the energy are conserved over time at the quadratic level. An obvious solution is to consider the equations of motion which are valid up to the second order in perturbations. This will be done in the next section.

\section{Charge and energy: correct calculations}
The equation of motion for the field $\varphi$ up to the terms quadratic in perturbations can easily be obtained and has the form
\begin{equation}\label{perteqsecond}
\omega^{2}\varphi-2i\omega\dot\varphi-\ddot\varphi+\Delta\varphi-U\varphi-S(\varphi+\varphi^{*})-
S\frac{1}{f}(\varphi^{2}+2\varphi^{*}\varphi)-J(\varphi+\varphi^{*})^{2}=0,
\end{equation}
where
\begin{equation}\label{Jdef}
J(r)=\frac{1}{2}\frac{d^{3}V}{d(\phi^{*}\phi)^{3}}\biggl|_{\phi^{*}\phi=f^{2}(r)}f^{3}(r).
\end{equation}
To simplify the subsequent calculations, it is convenient to get rid of the term $\partial_{i}\varphi^{*}\partial_{i}\varphi$ in \eqref{energygeneral}. This can be done by using equation of motion \eqref{perteqsecond}. Performing an integration by parts for the term with $\partial_{i}\varphi^{*}\partial_{i}\varphi$ in \eqref{energygeneral}, substituting equation of motion \eqref{perteqsecond}, and omitting the terms which are cubic in perturbations, we get
\begin{equation}\label{energypertreduced}
E_{p}=\omega Q_{p}+\int\left(i\omega(\dot\varphi^{*}\varphi-\varphi^{*}\dot\varphi)+\dot\varphi^{*}\dot\varphi-\frac{1}{2}\ddot\varphi^{*}\varphi-
\frac{1}{2}\varphi^{*}\ddot\varphi\right)d^{d}x.
\end{equation}

Below we will use the following form of the perturbation:
\begin{equation}\label{substgenintro}
\varphi(t,\vec x)=\alpha\psi_{1}(t,\vec x)+\alpha^{2}\psi_{2}(t,\vec x),
\end{equation}
where $\psi_{1}(t,\vec x)$ is supposed to satisfy linearized equation of motion \eqref{lineqgeneral}, whereas $\psi_{2}(t,\vec x)$ is supposed to be a correction coming from \eqref{perteqsecond}. In such a case, perturbation \eqref{substgenintro} solves nonlinear equation \eqref{perteqsecond} up to terms quadratic in $\alpha$. As in the previous section, the real parameter $\alpha\ll 1$ is introduced for the convenience --- with such a parameter we can suppose that the functions $\psi_{1}(t,\vec x)$ and $\psi_{2}(t,\vec x)$ are of the order of $f(r)$.

Now we are ready to consider specific examples.

\subsection{Time-independent modes}
First, let us consider perturbation of the form
\begin{equation}\label{substtind1}
\varphi(t,\vec x)=i\alpha f(r)+\alpha^{2}g(\vec x),
\end{equation}
where $g(\vec x)$ is some real function. Substituting \eqref{substtind1} into \eqref{perteqsecond} and isolating the terms $\sim\alpha$ and $\sim\alpha^{2}$, we get the equation
\begin{equation}
\hat L_{2}g+2Sg+Sf=0.
\end{equation}
Recalling that $\hat L_{2}f=0$, we get
\begin{equation}
g=-\frac{f}{2}.
\end{equation}
Substituting $\varphi(t,\vec x)=i\alpha f-\frac{\alpha^{2}}{2}f$ into \eqref{chargegeneral} and \eqref{energypertreduced} and keeping the terms $\sim\alpha$ and $\sim\alpha^{2}$, we get
\begin{equation}
Q_{p}=0,\qquad E_{p}=0.
\end{equation}
This is the expected result because
\begin{equation}
e^{i\alpha}f\approx f+i\alpha f-\frac{\alpha^{2}}{2}f.
\end{equation}
Meanwhile, it is clear that $i\alpha f-\frac{\alpha^{2}}{2}f$ is not a solution of linear equation \eqref{lineqgeneral}.

The second example, which can be considered here, is connected with the translational mode $\partial_{j}f(r)$,
\begin{equation}\label{substtrans}
\varphi(t,\vec x)=\alpha L\partial_{j}f(r)+\alpha^{2}g(\vec x),
\end{equation}
where $j=1,2,...,d$. Here $L$ is the parameter with the dimension of length that is introduced in order to keep $\alpha$ dimensionless. Substituting \eqref{substtrans} into \eqref{perteqsecond}, we get the following equation for the function $g(\vec x)$:
\begin{equation}\label{eqgtrans}
\hat L_{1}g+3S\frac{1}{f}(L\partial_{j}f)^{2}+4J(L\partial_{j}f)^{2}=0.
\end{equation}
Now let us recall that
\begin{equation}\label{trmodeeq}
\hat L_{1}\partial_{j}f=0,
\end{equation}
which can easily be obtained by differentiating equation $\hat L_{2}f=0$ with respect to $x^{j}$. Now let us differentiate Eq.~\eqref{trmodeeq} with respect to $x^{j}$. We get
\begin{equation}\label{transmodcorreq}
\hat L_{1}\partial_{j}^{2}f+2\Bigl(3S\frac{1}{f}(\partial_{j}f)^{2}+4J(\partial_{j}f)^{2}\Bigr)=0.
\end{equation}
Comparing this result with \eqref{eqgtrans}, we find that
\begin{equation}
g=\frac{L^{2}}{2}\partial_{j}^{2}f.
\end{equation}
Substituting
\begin{equation}
\varphi(t,\vec x)=\alpha L\partial_{j}f+\frac{\alpha^{2}L^{2}}{2}\partial_{j}^{2}f
\end{equation}
into \eqref{chargegeneral} and \eqref{energypertreduced}, we obtain
\begin{equation}
Q_{p}=0,\qquad E_{p}=0.
\end{equation}
This is also the expected result, because
\begin{equation}
f\left(x^{1},...,x^{j}+\alpha L,...,x^{d}\right)\approx f(r)+\alpha L\partial_{j}f(r)+\frac{\alpha^{2}L^{2}}{2}\partial_{j}^{2}f(r).
\end{equation}

\subsection{Modes corresponding to the Lorentz transformations}
For this mode, let us consider perturbation of the form
\begin{equation}\label{Lorentz2}
\varphi(t,\vec x)=v\left(t\partial_{j}f+i\omega x^{j}f\right)+v^{2}g(\vec x,t),
\end{equation}
where $g(\vec x)$ is some complex function. One can show that \eqref{Lorentz2} with
\begin{equation}\label{Lorentzg}
g(\vec x,t)=\frac{1}{2}x^{j}\partial_{j}f+\frac{1}{2}t^{2}\partial_{j}^{2}f-\frac{1}{2}\omega^{2}(x^{j})^{2}f+i\left(\omega tx^{j}\partial_{j}f+\frac{1}{2}\omega t f\right)
\end{equation}
satisfies Eq.~\eqref{perteqsecond} up to the quadratic order in $v$ (one should also use Eq.~\eqref{transmodcorreq} while performing the calculations). Note that there is no summation in $j$ in \eqref{Lorentzg}. With this solution, for the charge we get
\begin{equation}
Q_{p}=0,
\end{equation}
which is the expected result for the mode corresponding to a Q-ball moving with a constant speed.

Now we turn to the energy. Substituting \eqref{Lorentz2} into \eqref{energypertreduced}, we easily get
\begin{equation}\label{Lorentzen}
E_{p}=v^{2}\int\left(\omega^{2}f^{2}+\partial_{j}f\partial_{j}f\right)d^{d}x.
\end{equation}
Using the spherical symmetry of the Q-ball profile $f$, the integral in \eqref{Lorentzen} can be rewritten as
\begin{equation}
\int\left(\omega^{2}f^{2}+\partial_{j}f\partial_{j}f\right)d^{d}x=\frac{1}{2}\left(\omega Q_{0}+\frac{2}{d}\int \sum_{l=1}^{d}\partial_{l}f\partial_{l}f\, d^{d}x\right),
\end{equation}
where $Q_{0}$ is Q-ball charge \eqref{qballcharge}. Since
\begin{equation}
\omega Q_{0}+\frac{2}{d}\int\sum_{l=1}^{d}\partial_{l}f\partial_{l}f\,d^{d}x=E_{0},
\end{equation}
where $E_{0}$ is the Q-ball energy (the proof can be found in Appendix~A of \cite{Gulamov:2013ema}), for the energy we obtain
\begin{equation}
E_{p}=E_{0}\frac{v^{2}}{2},
\end{equation}
which is the kinetic energy of the moving Q-ball in the nonrelativistic approximation. This is the expected result, because
\begin{align}\nonumber
\phi(t,\vec x)=e^{i\omega\left(\frac{t+vx^{j}}{\sqrt{1-v^{2}}}\right)}f\left(x^{1},...,\frac{x^{j}+vt}{\sqrt{1-v^{2}}},...,x^{d}\right)\\\approx
e^{i\omega t}f(r)+e^{i\omega t}\left(v\left(t\partial_{j}f+i\omega x^{j}f\right)+v^{2}g(\vec x,t)\right),
\end{align}
where $g(\vec x,t)$ is defined by \eqref{Lorentzg}.

\subsection{Instability modes}
\label{subsdecmodes2}
For the instability mode, we consider perturbation of the form
\begin{equation}\label{substdeccorr}
\varphi(t,\vec x)=\alpha e^{\gamma t}\Bigl(u(\vec x)+iv(\vec x)\Bigr)+\alpha^{2}e^{2\gamma t}\Bigl(p(\vec x)+iq(\vec x)\Bigr),
\end{equation}
where $\gamma$ is real and $\gamma\neq 0$; $u(\vec x), v(\vec x), p(\vec x), q(\vec x)$ are real functions. Substituting \eqref{substdeccorr} into Eq.~\eqref{perteqsecond}, keeping the terms $\sim\alpha$ and $\sim\alpha^{2}$ and isolating the purely real and purely imaginary terms with different dependences on time, we get Eqs.~\eqref{equv1} and \eqref{equv2} and the following equations for the functions $p$, $q$:
\begin{align}
-\hat L_{1}p+4\omega\gamma q-4\gamma^{2}p&=S\frac{1}{f}(3u^{2}+v^{2})+4Ju^{2},\\\label{pquveq}
-\hat L_{2}q-4\omega\gamma p-4\gamma^{2}q&=2S\frac{1}{f}uv.
\end{align}
Substituting \eqref{substdeccorr} into \eqref{chargegeneral} and retaining the terms $\sim\alpha$ and $\sim\alpha^{2}$, we get
\begin{equation}\label{chargedecaycorr}
Q_{p}=\alpha e^{\gamma t}\int(4\omega fu+2\gamma fv)d^{d}x+\alpha^{2}e^{2\gamma t}\int\Bigl(4\omega f p+4\gamma f q+2\omega(u^{2}+v^{2})\Bigr)d^{d}x.
\end{equation}
The term $\sim\alpha$ vanishes because of \eqref{linvanishdecay}, so we are left with only the term $\sim\alpha^{2}$.

Now let us multiply Eq.~\eqref{pquveq} by $f$ and integrate over the space. We get
\begin{equation}\label{auxdecay1}
\int\left(4\omega f p+4\gamma f q+\frac{2}{\gamma}Suv\right)d^{d}x=0.
\end{equation}
Then we consider the relation $\int d^{d}x(v\hat L_{1}u-u\hat L_{2}v)=2\int d^{d}x\,Suv$, which follows from definition of the operators $\hat L_{1}$ and $\hat L_{2}$, \eqref{L1eq} and \eqref{L2eq}. Using Eqs.~\eqref{equv1} and \eqref{equv2}, we get for this relation
\begin{equation}\label{auxdecay2}
2\int Suv\,d^{d}x=2\omega\gamma\int(u^{2}+v^{2})d^{d}x.
\end{equation}
Combining \eqref{auxdecay1} and \eqref{auxdecay2}, we obtain
\begin{equation}
\int\left(4\omega f p+4\gamma f q+2\omega(u^{2}+v^{2})\right)d^{d}x=0,
\end{equation}
which means that the term $\sim\alpha^{2}$ in \eqref{chargedecaycorr} also vanishes. Using \eqref{energypertreduced}, finally we get
\begin{equation}\label{decmodechen}
Q_{p}=0, \qquad E_{p}=0,
\end{equation}
which is the expected result for the instability mode.

\subsection{Oscillation modes}\label{subsoscmodes2}
Finally, let us turn to the oscillation mode and consider perturbation of the form
\begin{equation}\label{substosccorr}
\varphi(t,\vec x)=\alpha\left(a(\vec x)e^{i\gamma t}+b(\vec x)e^{-i\gamma t}\right)+\alpha^{2}\left(m(\vec x)+j(\vec x)e^{2i\gamma t}+w(\vec x)e^{-2i\gamma t}\right),
\end{equation}
where $\gamma$ is real and $\gamma>0$; $a(\vec x)$, $b(\vec x)$, $m(\vec x)$, $j(\vec x)$ and $w(\vec x)$ are complex functions. From the very beginning, it is convenient to use notations \eqref{notatosc} and the notations
\begin{align}
\chi_{+}(\vec x)=m(\vec x)+m^{*}(\vec x),\qquad \chi_{-}(\vec x)=m(\vec x)-m^{*}(\vec x),\\
\psi_{+}(\vec x)=j(\vec x)+w^{*}(\vec x),\qquad \psi_{-}(\vec x)=j(\vec x)-w^{*}(\vec x).
\end{align}
Substituting \eqref{substosccorr} into Eq.~\eqref{perteqsecond}, keeping the terms $\sim\alpha$ and $\sim\alpha^{2}$, and isolating the terms with different dependences on time, we get Eqs.~\eqref{lineq1} and \eqref{lineq2} and the following equations for $\chi_{+}$, $\chi_{-}$, $\psi_{+}$, and $\psi_{-}$:
\begin{align}\label{chipluseqsmode}
-\hat L_{1}\chi_{+}&=S\frac{1}{f}(3\xi^{*}\xi+\eta^{*}\eta)+4J\xi^{*}\xi,\\
-\hat L_{2}\chi_{-}&=S\frac{1}{f}(\xi^{*}\eta-\eta^{*}\xi),\\
-\hat L_{1}\psi_{+}+4\omega\gamma\psi_{-}+4\gamma^{2}\psi_{+}&=S\frac{1}{f}\left(\frac{3}{2}\xi^{2}-\frac{1}{2}\eta^{2}\right)+2J\xi^{2},\\\label{singpartpsi2}
-\hat L_{2}\psi_{-}+4\omega\gamma\psi_{+}+4\gamma^{2}\psi_{-}&=S\frac{1}{f}\xi\eta.
\end{align}
Now we substitute \eqref{substosccorr} into \eqref{chargegeneral} and retain the terms $\sim\alpha$ and $\sim\alpha^{2}$. The result looks as follows:
\begin{align}\nonumber
Q_{p}=\alpha\int\Bigl(e^{i\gamma t}(2\omega f\xi+\gamma f\eta)+e^{-i\gamma t}(2\omega f\xi^{*}+\gamma f\eta^{*})\Bigr)d^{d}x\\\nonumber+
\alpha^{2}\int d^{d}x\Bigl(2\omega f\chi_{+}+\omega(\xi^{*}\xi+\eta^{*}\eta)+\gamma(\xi^{*}\eta+\eta^{*}\xi)\\\label{Qonemodeprelim}
+e^{2i\gamma t}\frac{1}{2}\bigl(4\omega f\psi_{+}+4\gamma f\psi_{-}+\omega(\xi^{2}-\eta^{2})\bigr)+e^{-2i\gamma t}\frac{1}{2}\bigl(4\omega f\psi_{+}^{*}+4\gamma f\psi_{-}^{*}+\omega({\xi^{*}}^{2}-{\eta^{*}}^{2})\bigr)\Bigr).
\end{align}
It is possible to show that
\begin{equation}\label{appAeq}
\int\Bigl(4\omega f\psi_{+}+4\gamma f\psi_{-}+\omega(\xi^{2}-\eta^{2})\Bigr)d^{d}x=0;
\end{equation}
see Appendix~B for details. Together with \eqref{osclinterm}, it means that the terms $\sim e^{\pm i\gamma t}$ and $\sim e^{\pm 2i\gamma t}$ vanish from \eqref{Qonemodeprelim}. Thus, for the charge we get
\begin{equation}\label{chargesingpart0}
Q_{p}=\alpha^{2}\int\Bigl(2\omega f\chi_{+}+\omega(\xi^{*}\xi+\eta^{*}\eta)+\gamma(\xi^{*}\eta+\eta^{*}\xi)\Bigr)d^{d}x.
\end{equation}
Substituting \eqref{substosccorr} into \eqref{energypertreduced}, we easily obtain
\begin{equation}\label{energysingpart0}
E_{p}=\alpha^{2}\int\Bigl(2\omega^{2}f\chi_{+}+(\omega^{2}+\gamma^{2})(\xi^{*}\xi+\eta^{*}\eta)+2\omega\gamma(\xi^{*}\eta+\eta^{*}\xi)\Bigr)d^{d}x.
\end{equation}
Note that the term with $\chi_{+}$ in \eqref{energysingpart0}, i.e., the nonlinear correction, comes only from the term $\omega Q_{p}$ of \eqref{energypertreduced}. One can see that this nonlinear correction is equal to zero for $\omega=0$. Thus, for the static background one can use the formulas without the nonlinear corrections (compare, say, \eqref{chargenoncons} with $\omega=0$ and \eqref{chargesingpart0} with $\omega=0$).

To get the charge and the energy of the mode, it is necessary to find $\chi_{+}$. However, if one knows not only the Q-ball profile $f(r)$ for a given value of the frequency $\omega$, but also the forms of Q-ball solutions in the vicinity of this frequency $\omega$ (or if there exists the analytic solution $f(r,\omega)$), it is possible to simplify the task of calculating $Q_{p}$ and $E_{p}$ considerably by using the following trick:
let us multiply Eq.~\eqref{chipluseqsmode} by $\frac{df}{d\omega}$, integrate the result over the space, and use Eq.~\eqref{dfdomegaL1}. We obtain
\begin{equation}
\int 2\omega f\chi_{+}d^{d}x=\int \frac{df}{d\omega}\hat L_{1}\chi_{+}d^{d}x=-\int \frac{df}{d\omega}\left(S\frac{1}{f}(3\xi^{*}\xi+\eta^{*}\eta)+4J\xi^{*}\xi\right)d^{d}x,
\end{equation}
leading to
\begin{align}\label{chargesinglepart}
&Q_{p}=\alpha^{2}\int \Bigl(\omega(\xi^{*}\xi+\eta^{*}\eta)+\gamma(\xi^{*}\eta+\eta^{*}\xi)-\frac{1}{f}\frac{df}{d\omega}S(3\xi^{*}\xi+\eta^{*}\eta)-4\frac{df}{d\omega}J\xi^{*}\xi\Bigr)d^{d}x.
\\\label{energysinglepart}
&E_{p}=\alpha^{2}\int\Bigl((\omega^{2}+\gamma^{2})(\xi^{*}\xi+\eta^{*}\eta)+2\omega\gamma(\xi^{*}\eta+\eta^{*}\xi)
-\frac{\omega}{f}\frac{df}{d\omega}S(3\xi^{*}\xi+\eta^{*}\eta)-4\omega\frac{df}{d\omega}J\xi^{*}\xi\Bigr)d^{d}x.
\end{align}
A remarkable feature of formulas \eqref{chargesinglepart} and \eqref{energysinglepart} is that in order to correctly calculate the charge and the energy of the oscillation mode, it is sufficient to have only the solutions to linearized equations of motion (together with $\frac{df}{d\omega}$), whereas all the nonlinear corrections turn out be taken into account automatically in \eqref{chargesinglepart} and \eqref{energysinglepart}. Note that $Q_{p}$ and $E_{p}$ are finite for the modes from a discrete spectrum.

We see that, contrary to formula \eqref{chargenoncons}, here the charge $Q_{p}$ is conserved over time, as well as the energy $E_{p}$. Moreover, even if we average \eqref{chargenoncons} over the ``period of oscillations'' $\frac{2\pi}{\gamma}$ (here I do not take into account the overall factor $e^{i\omega t}$), we will not get the correct result for the charge of the oscillation mode (because of the absence of the necessary term with $\chi_{+}$).

In the general case, $Q_{p}$ and $E_{p}$ defined by \eqref{chargesinglepart} and \eqref{energysinglepart} can be negative or positive. A possible negativity of $E_{p}$ is not strange --- $E_{p}$ is not the absolute energy; it is just the relative energy of the mode with respect to the Q-ball energy. Namely, one has to extract the energy $-E_{p}>0$ from the Q-ball (and at the same time to extract the charge $-Q_{p}$ from the Q-ball) to produce this mode. However, one may think that it is energetically favorable for the Q-ball to drop the charge $-Q_{p}$ and the energy $-E_{p}$, say, by emitting some particles. The latter may look like a potential source of classical or quantum instability. Let us discuss this problem in more detail.

First, from \eqref{chargesinglepart} and \eqref{energysinglepart} we see that
\begin{equation}\label{negEpeq1}
E_{p}=\omega Q_{p}+\alpha^{2}\int\Bigl(\gamma^{2}(\xi^{*}\xi+\eta^{*}\eta)+\omega\gamma(\xi^{*}\eta+\eta^{*}\xi)\Bigr)d^{d}x.
\end{equation}
Now let us take Eq.~\eqref{lineq2}, multiply it by $\eta^{*}$ and integrate the result over the space. We obtain
\begin{equation}\label{xietaetaxi}
\int\Bigl(\eta^{*}\hat L_{2}\eta-\gamma^{2}\eta^{*}\eta\Bigr)d^{d}x=2\omega\gamma\int\eta^{*}\xi\, d^{d}x.
\end{equation}
Combining relation \eqref{xietaetaxi} with its complex conjugate, we get
\begin{equation}
\int\Bigl(\eta^{*}\hat L_{2}\eta-\gamma^{2}\eta^{*}\eta\Bigr)d^{d}x=\omega\gamma\int (\eta^{*}\xi+\xi^{*}\eta)d^{d}x.
\end{equation}
Using the latter relation, for \eqref{negEpeq1} we obtain
\begin{equation}\label{negEpeq2}
E_{p}=\omega Q_{p}+\alpha^{2}\int\Bigl(\gamma^{2}\xi^{*}\xi+\eta^{*}\hat L_{2}\eta\Bigr)d^{d}x.
\end{equation}
The inequality $\int d^{d}x\,\eta^{*}\hat L_{2}\eta\ge 0$ (see \eqref{L2positive}) results in
\begin{equation}\label{negEpeq3}
\int\Bigl(\gamma^{2}\xi^{*}\xi+\eta^{*}\hat L_{2}\eta\Bigr)d^{d}x>0
\end{equation}
for $\gamma\neq 0$, $\xi\not\equiv 0$ (if $\xi\equiv 0$, then, according to Eq.~\eqref{lineq1}, $\eta\equiv 0$). Inequality \eqref{negEpeq3} implies that $E_{p}$ can be negative only if $\omega Q_{p}<0$.

However, in such a case there is a more energetically favorable process than the one mentioned above. Namely, if the Q-ball simply changes its frequency $\omega$ by dropping (somehow) the charge $-Q_{p}$, then the energy loss of the initial Q-ball will be larger than the one in the case of the ``excited state'' with $E_{p}<0$:
\begin{equation}\label{negEpeq4}
|\triangle E_{0}|\approx|\omega Q_{p}|>|\omega Q_{p}|-\int\Bigl(\gamma^{2}\xi^{*}\xi+\eta^{*}\hat L_{2}\eta\Bigr)d^{d}x=|E_{p}|.
\end{equation}
In deriving this inequality, it is necessary to use the relation
\begin{equation}\label{dEdQ}
\frac{dE_{0}}{dQ_{0}}=\omega,
\end{equation}
which holds for any Q-ball.

Now we have two possibilities. First, let there exist only the scalar excitations of the field $\phi$ against the vacuum solution $\phi\equiv 0$, forming the corresponding scalar particles of the theory. The mass of these particles $M=\sqrt{\frac{dV}{d(\phi^{*}\phi)}\bigl|_{\phi^{*}\phi=0}}$ is such that $|\omega|<M$; thus, both processes with the emitting of such scalar particles are energetically forbidden. Of course, if $E_{0}<MQ_{0}$, the quantum decay of the Q-ball into scalar particles is possible, but it still cannot go through the excited state with $E_{p}<0$. Second, if there exist some additional particles of mass $m<|\omega|$, which interact with the Q-ball and carry the same U(1) charge as the Q-ball does, both processes are not forbidden, but the process ending up with the new Q-ball of energy $E_{0}+\triangle E_{0}\approx E_{0}+\omega Q_{p}<E_{0}$ is more energetically favorable than the process ending up with the excited Q-ball with $E_{p}<0$. All these reasonings show that the existence of the oscillation mode with $E_{p}<0$ is not dangerous and does not lead to any additional instability, at least at the quantum level. As for the classical stability, one should not expect that the existence of such modes can somehow affect the classical stability of Q-balls, which can easily be seen from the derivation of the classical stability criterion \cite{Friedberg:1976me,LeePang} based on the use of linearized equations of motion for perturbations \cite{Panin:2016ooo}.\footnote{An explicit example of the mode with $E_{p}<0$ for classically stable Q-balls in the model with logarithmic scalar field potential will be presented in Section~\ref{secexpexamp}.} However, we will also discuss this problem at the end of the next section.

\section{Additivity of charge and energy for oscillation modes}
In the previous section the formulas were obtained for the charge and the energy (in the quadratic order in perturbations) of a single mode forming the perturbation against a Q-ball. But what if we have many modes in the perturbation? The main question is whether the additivity property is valid for the charge and the energy. The naive expectation is that it is not valid because of the nonlinear equation of motion for the perturbations. However, this naive expectation is not correct.

Suppose we have a classically stable Q-ball (in order not to deal with the instability modes of form \eqref{substdecayincorrect}). Let us take the following solution to linearized equation of motion \eqref{lineqgeneral}:
\begin{equation}\label{pertlin}
\varphi_{lin}(t,\vec x)=\alpha\sum_{n}\left(a_{n}(\vec x)e^{i\gamma_{n} t}+b_{n}(\vec x)e^{-i\gamma_{n} t}\right),
\end{equation}
where $\gamma_{n}$ are real, $\gamma_{n}>0$ and $\gamma_{n}\neq\gamma_{k}$ for $n\neq k$. Without loss of generality, I take one and the same $\alpha$ for all modes. Also, for simplicity, I take only the modes from the discrete part of the spectrum. However, the modes from the continuous part of the spectrum, if they exist, can easily be taken into account: the simplest way to do it is to put the system into a ``box'' of a finite size.

The perturbation with nonlinear corrections takes the form
\begin{align}\nonumber
\varphi(t,\vec x)=\alpha\sum_{n}\left(a_{n}(\vec x)e^{i\gamma_{n} t}+b_{n}(\vec x)e^{-i\gamma_{n} t}\right)+\alpha^{2}m(\vec x)+
\alpha^{2}\sum_{n}\left(j_{n}(\vec x)e^{2i\gamma_{n} t}+w_{n}(\vec x)e^{-2i\gamma_{n}t}\right)\\\label{substoscsup}
+\alpha^{2}\sum_{\substack{n,k\\n<k}}\left(p_{1,nk}(\vec x)e^{i(\gamma_{n}+\gamma_{k})t}+p_{2,nk}(\vec x)e^{-i(\gamma_{n}+\gamma_{k})t}+q_{1,nk}(\vec x)e^{i(\gamma_{n}-\gamma_{k})t}+q_{2,nk}(\vec x)e^{-i(\gamma_{n}-\gamma_{k})t}\right),
\end{align}
where $a_{n}(\vec x)$, $b_{n}(\vec x)$, $m(\vec x)$, $j_{n}(\vec x)$, $w_{n}(\vec x)$, $p_{1,nk}(\vec x)$, $p_{2,nk}(\vec x)$, $q_{1,nk}(\vec x)$, and $q_{2,nk}(\vec x)$ are complex functions. It is also convenient to use the notations
\begin{align}
\xi_{n}(\vec x)=a_{n}(\vec x)+b_{n}^{*}(\vec x),&\qquad \eta_{n}(\vec x)=a_{n}(\vec x)-b_{n}^{*}(\vec x),\\
\chi_{+}(\vec x)=m(\vec x)+m^{*}(\vec x),&\qquad \chi_{-}(\vec x)=m(\vec x)-m^{*}(\vec x),\\
\psi_{+,n}(\vec x)=j_{n}(\vec x)+w_{n}^{*}(\vec x),&\qquad \psi_{-,n}(\vec x)=j_{n}(\vec x)-w_{n}^{*}(\vec x),\\
\varrho_{+,nk}(\vec x)=p_{1,nk}(\vec x)+p_{2,nk}^{*}(\vec x),&\qquad \varrho_{-,nk}(\vec x)=p_{1,nk}(\vec x)-p_{2,nk}^{*}(\vec x),\\
\theta_{+,nk}(\vec x)=q_{1,nk}(\vec x)+q_{2,nk}^{*}(\vec x),&\qquad \theta_{-,nk}(\vec x)=q_{1,nk}(\vec x)-q_{2,nk}^{*}(\vec x).
\end{align}
Substituting perturbation \eqref{substoscsup} into Eq.~\eqref{perteqsecond}, keeping the terms $\sim\alpha$ and $\sim\alpha^{2}$, and isolating the terms with different dependences on time, we get the following set of equations:
\begin{align}\label{eqxin}
-\hat L_{1}\xi_{n}+2\omega\gamma_{n}\eta_{n}+\gamma_{n}^{2}\xi_{n}&=0,\\\label{eqetan}
-\hat L_{2}\eta_{n}+2\omega\gamma_{n}\xi_{n}+\gamma_{n}^{2}\eta_{n}&=0,\\\label{chipluseq}
-\hat L_{1}\chi_{+}&=S\frac{1}{f}\sum_{n}\left(3\xi_{n}^{*}\xi_{n}+\eta_{n}^{*}\eta_{n}\right)+4J\sum_{n}\xi_{n}^{*}\xi_{n},\\
-\hat L_{2}\chi_{-}&=S\frac{1}{f}\sum_{n}\left(\xi_{n}^{*}\eta_{n}-\eta_{n}^{*}\xi_{n}\right),
\end{align}
\begin{align}
-\hat L_{1}\psi_{+,n}+4\omega\gamma_{n}\psi_{-,n}+4\gamma_{n}^{2}\psi_{+,n}&=S\frac{1}{f}\left(\frac{3}{2}\xi_{n}^{2}-\frac{1}{2}\eta_{n}^{2}\right)+2J\xi_{n}^{2},\\
-\hat L_{2}\psi_{-,n}+4\omega\gamma_{n}\psi_{+,n}+4\gamma_{n}^{2}\psi_{-,n}&=S\frac{1}{f}\xi_{n}\eta_{n},\\
-\hat L_{1}\varrho_{+,nk}+2\omega(\gamma_{n}+\gamma_{k})\varrho_{-,nk}+(\gamma_{n}+\gamma_{k})^{2}\varrho_{+,nk}&=
S\frac{1}{f}\left(3\xi_{n}\xi_{k}-\eta_{n}\eta_{k}\right)+4J\xi_{n}\xi_{k},\\\label{eqvarrho}
-\hat L_{2}\varrho_{-,nk}+2\omega(\gamma_{n}+\gamma_{k})\varrho_{+,nk}+(\gamma_{n}+\gamma_{k})^{2}\varrho_{-,nk}&=
S\frac{1}{f}\left(\xi_{n}\eta_{k}+\eta_{n}\xi_{k}\right),\\
-\hat L_{1}\theta_{+,nk}+2\omega(\gamma_{n}-\gamma_{k})\theta_{-,nk}+(\gamma_{n}-\gamma_{k})^{2}\theta_{+,nk}&=
S\frac{1}{f}\left(3\xi_{n}\xi_{k}^{*}+\eta_{n}\eta_{k}^{*}\right)+4J\xi_{n}\xi_{k}^{*},\\\label{eqtheta}
-\hat L_{2}\theta_{-,nk}+2\omega(\gamma_{n}-\gamma_{k})\theta_{+,nk}+(\gamma_{n}-\gamma_{k})^{2}\theta_{-,nk}&=
S\frac{1}{f}\left(\eta_{n}\xi_{k}^{*}-\xi_{n}\eta_{k}^{*}\right).
\end{align}
Now let us substitute \eqref{substoscsup} into \eqref{chargegeneral} and keep the terms $\sim\alpha$ and $\sim\alpha^{2}$. In full analogy with the case of a single mode, the terms linear in $\alpha$ (which are $\sim e^{\pm i\gamma_{n} t}$) vanish because of \eqref{osclinterm}, whereas the terms $\sim e^{\pm 2i\gamma_{n} t}$ vanish because of \eqref{appAeq}. Thus, we obtain
\begin{align}\nonumber
Q_{p}=\alpha^{2}\int\left(2\omega f\chi_{+}+\sum_{n}\Bigl(\omega(\xi_{n}\xi_{n}^{*}+\eta_{n}\eta_{n}^{*})+\gamma_{n}(\xi_{n}\eta_{n}^{*}+\eta_{n}\xi_{n}^{*})\Bigr)\right)d^{d}x+
\int d^{d}x\sum_{\substack{n,k\\n<k}}\alpha^{2}\\\nonumber\times\left[e^{i(\gamma_{n}+\gamma_{k})t}\left(2\omega f\varrho_{+,nk}+(\gamma_{n}+\gamma_{k})f\varrho_{-,nk}
+\omega(\xi_{n}\xi_{k}-\eta_{n}\eta_{k})+\frac{1}{2}(\gamma_{n}-\gamma_{k})(\xi_{k}\eta_{n}-\xi_{n}\eta_{k})\right)\right.\\\label{chargeprelim}
\left.+e^{i(\gamma_{n}-\gamma_{k})t}\left(
2\omega f\theta_{+,nk}+(\gamma_{n}-\gamma_{k})f\theta_{-,nk}
+\omega(\xi_{n}\xi_{k}^{*}+\eta_{n}\eta_{k}^{*})+\frac{1}{2}(\gamma_{n}+\gamma_{k})(\xi_{k}^{*}\eta_{n}+\xi_{n}\eta_{k}^{*})
\right)+\textrm{c.c.}\right].
\end{align}
It is possible to show that
\begin{align}\label{appBeq1}
\int\left(2\omega f\varrho_{+,nk}+(\gamma_{n}+\gamma_{k})f\varrho_{-,nk}+
\omega(\xi_{n}\xi_{k}-\eta_{n}\eta_{k})+\frac{1}{2}(\gamma_{n}-\gamma_{k})(\xi_{k}\eta_{n}-\xi_{n}\eta_{k})\right)d^{d}x=0,\\\label{appBeq2}
\int\left(2\omega f\theta_{+,nk}+(\gamma_{n}-\gamma_{k})f\theta_{-,nk}+\omega(\eta_{n}\eta_{k}^{*}+\xi_{n}\xi_{k}^{*})+
\frac{1}{2}(\gamma_{n}+\gamma_{k})(\xi_{k}^{*}\eta_{n}+\xi_{n}\eta_{k}^{*})\right)d^{d}x=0;
\end{align}
see Appendix~C for details. It means that the terms $\sim e^{\pm i(\gamma_{n}+\gamma_{k})t}$ and $\sim e^{\pm i(\gamma_{n}-\gamma_{k})t}$ also vanish from \eqref{chargeprelim}, and we arrive at
\begin{equation}\label{chargesuper0}
Q_{p}=\alpha^{2}\int\left(2\omega f\chi_{+}+\sum_{n}\Bigl(\omega(\xi_{n}\xi_{n}^{*}+\eta_{n}\eta_{n}^{*})+\gamma_{n}(\xi_{n}\eta_{n}^{*}+\eta_{n}\xi_{n}^{*})\Bigr)\right)d^{d}x.
\end{equation}
It is clear that, according to \eqref{chipluseq}, one can represent $\chi_{+}$ as
\begin{equation}
\chi_{+}=\sum_{n}\chi_{+}^{(n)},
\end{equation}
where each $\chi_{+}^{(n)}$ satisfies the equation
\begin{equation}\label{chiplusneq}
-\hat L_{1}\chi_{+}^{(n)}=S\frac{1}{f}\left(3\xi_{n}^{*}\xi_{n}+\eta_{n}^{*}\eta_{n}\right)+4J\xi_{n}^{*}\xi_{n}.
\end{equation}
Then, $Q_{p}$ takes the form
\begin{align}\nonumber
Q_{p}=\alpha^{2}\sum_{n}\int\Bigl(2\omega f\chi_{+}^{(n)}+\omega(\xi_{n}\xi_{n}^{*}+\eta_{n}\eta_{n}^{*})+\gamma_{n}(\xi_{n}\eta_{n}^{*}+\eta_{n}\xi_{n}^{*})\Bigr)d^{d}x\\ \label{chargesuper1}
=\alpha^{2}\sum_{n}\int\left(\omega(\xi_{n}\xi_{n}^{*}+\eta_{n}\eta_{n}^{*})+\gamma_{n}(\xi_{n}\eta_{n}^{*}+\eta_{n}\xi_{n}^{*})-
\frac{1}{f}\frac{df}{d\omega}S(3\xi_{n}\xi_{n}^{*}+\eta_{n}\eta_{n}^{*})-4\frac{df}{d\omega}J\xi_{n}\xi_{n}^{*}\right)d^{d}x.
\end{align}

Substituting \eqref{substoscsup} into \eqref{energypertreduced} and keeping the terms $\sim\alpha$ and $\sim\alpha^{2}$, for the energy we obtain
\begin{align}\nonumber
E_{p}=\omega Q_{p}+\alpha^{2}\int \sum_{n}\Bigl(\gamma_{n}^{2}(\xi_{n}^{*}\xi_{n}+\eta_{n}^{*}\eta_{n})+\omega\gamma_{n}(\xi_{n}^{*}\eta_{n}+\eta_{n}^{*}\xi_{n})\Bigr)d^{d}x\\\nonumber
+\frac{\alpha^{2}}{4}\int\sum_{\substack{n,k\\n<k}}\Biggl(
e^{i(\gamma_{n}+\gamma_{k})t}(\gamma_{n}-\gamma_{k})\Bigl(2\omega(\eta_{n}\xi_{k}-\xi_{n}\eta_{k})+(\gamma_{n}-\gamma_{k})(\xi_{n}\xi_{k}-\eta_{n}\eta_{k})\Bigr)\\\label{energyprelim}
+e^{i(\gamma_{n}-\gamma_{k})t}(\gamma_{n}+\gamma_{k})\Bigl(2\omega(\eta_{n}\xi_{k}^{*}+\xi_{n}\eta_{k}^{*})+(\gamma_{n}+\gamma_{k})
(\xi_{n}\xi_{k}^{*}+\eta_{n}\eta_{k}^{*})\Bigr)+\textrm{c.c.}\Biggr)d^{d}x.
\end{align}
It is possible to show that
\begin{align}\label{appCeq1}
\int\left(2\omega(\eta_{n}\xi_{k}-\eta_{k}\xi_{n})+(\gamma_{n}-\gamma_{k})(\xi_{n}\xi_{k}-\eta_{n}\eta_{k})\right)d^{d}x=0,\\\label{appCeq2}
\int\left(2\omega(\eta_{n}\xi_{k}^{*}+\xi_{n}\eta_{k}^{*})+(\gamma_{n}+\gamma_{k})
(\xi_{n}\xi_{k}^{*}+\eta_{n}\eta_{k}^{*})\right)d^{d}x=0;
\end{align}
see Appendix~D for details. It means that the terms $\sim e^{\pm i(\gamma_{n}+\gamma_{k})t}$ and $\sim e^{\pm i(\gamma_{n}-\gamma_{k})t}$ vanish from \eqref{energyprelim}, and we arrive at
\begin{equation}\label{energysuper1}
E_{p}=\omega Q_{p}+\alpha^{2}\sum_{n}\int \Bigl(\gamma_{n}^{2}(\xi_{n}^{*}\xi_{n}+\eta_{n}^{*}\eta_{n})+\omega\gamma_{n}(\xi_{n}^{*}\eta_{n}+\eta_{n}^{*}\xi_{n})\Bigr)d^{d}x.
\end{equation}

We see that the total charge of the perturbation is just the sum of the charges of the modes. It means that the additivity property is valid for the charge, though we used nonlinear equation of motion \eqref{perteqsecond} for the perturbation. That is,
\begin{equation}\label{chargesuperpos}
Q_{p}=\sum_{n}Q_{p}^{(n)}[\xi_{n},\eta_{n},\gamma_{n}],
\end{equation}
where each $Q_{p}^{(n)}[\xi_{n},\eta_{n},\gamma_{n}]$ can be calculated using formula \eqref{chargesinglepart}. Analogously, we can write
\begin{equation}\label{energysuperpos}
E_{p}=\sum_{n}E_{p}^{(n)}[\xi_{n},\eta_{n},\gamma_{n}],
\end{equation}
where each $E_{p}^{(n)}[\xi_{n},\eta_{n},\gamma_{n}]$ can be calculated using formula \eqref{energysinglepart}. The additivity property is valid for the energy too.

What if we have two modes such that $\gamma_{i}=\gamma_{j}=\gamma$ for $\xi_{i}\not\equiv c\xi_{j}$ and $\eta_{i}\not\equiv c\eta_{j}$, where $c$ is a constant, i.e., different modes of the same frequency? In this case one gets in \eqref{chargesuper0} and \eqref{energysuper1} $\xi_{i}+\xi_{j}$ instead of a single $\xi_{n}$ and $\eta_{i}+\eta_{j}$ instead of a single $\eta_{n}$. There is a good reason to believe that the corresponding overlap integrals in \eqref{chargesuper0} and \eqref{energysuper1} are equal to zero, i.e.,
\begin{align}\label{supeqfreq1}
Q_{p}^{(i+j)}[\xi_{i}+\xi_{j},\eta_{i}+\eta_{j},\gamma]=Q_{p}^{(i)}[\xi_{i},\eta_{i},\gamma]+Q_{p}^{(j)}[\xi_{j},\eta_{j},\gamma],\\\label{supeqfreq2} E_{p}^{(i+j)}[\xi_{i}+\xi_{j},\eta_{i}+\eta_{j},\gamma]=E_{p}^{(i)}[\xi_{i},\eta_{i},\gamma]+E_{p}^{(j)}[\xi_{j},\eta_{j},\gamma].
\end{align}
Indeed, let us use the standard trick and modify ``by hands'' the function $S(r)\to S_{(\epsilon)}(r)=S(r)+\epsilon\,\delta S(r)$. Here $\delta S(r)$ is chosen in such a way that it removes the degeneracy of the modes
\begin{equation}
\gamma_{i}\to \gamma_{i}^{(\epsilon)}=\gamma+\epsilon\,\delta\gamma_{i},\qquad \gamma_{j}\to \gamma_{j}^{(\epsilon)}=\gamma+\epsilon\,\delta\gamma_{j},
\end{equation}
where $\delta\gamma_{i}\neq \delta\gamma_{j}$. Of course, the functions $\xi_{i,j}$ and $\eta_{i,j}$ (as well as the functions $\chi_{+}^{(i,j)}$, which are expressed through $\xi_{i,j}$ and $\eta_{i,j}$ by means of Eq.~\eqref{chiplusneq}) also turn out to be modified as
\begin{equation}
\xi_{i,j}\to \xi_{i,j}^{(\epsilon)}=\xi_{i,j}+\epsilon\,\delta\xi_{i,j},\qquad \eta_{i,j}\to \eta_{i,j}^{(\epsilon)}=\eta_{i,j}+\epsilon\,\delta\eta_{i,j}.
\end{equation}
In this case $\gamma_{i}^{(\epsilon)}\neq\gamma_{j}^{(\epsilon)}$ and, according to the results presented above, the terms $\sim e^{\pm i(\gamma_{i}^{(\epsilon)}\pm \gamma_{j}^{(\epsilon)})t}$ vanish, so we are left with
\begin{equation}
Q_{p}^{(i)}[\xi_{i}^{(\epsilon)},\eta_{i}^{(\epsilon)},\gamma_{i}^{(\epsilon)}]+Q_{p}^{(j)}[\xi_{j}^{(\epsilon)},\eta_{j}^{(\epsilon)},\gamma_{j}^{(\epsilon)}]
\end{equation}
for the charge and
\begin{equation}
E_{p}^{(i)}[\xi_{i}^{(\epsilon)},\eta_{i}^{(\epsilon)},\gamma_{i}^{(\epsilon)}]+E_{p}^{(j)}[\xi_{j}^{(\epsilon)},\eta_{j}^{(\epsilon)},\gamma_{j}^{(\epsilon)}]
\end{equation}
for the energy.\footnote{The modification of $S(r)$ may even remove the degeneracy of the time-independent modes $if$ and $\partial_{i}f$. Meanwhile, since the operator $\hat L_{2}$ remains intact and $\hat L_{2}f=0$, the results of Appendices~C and D remain valid and we still can obtain formulas \eqref{chargesuper0} (but not \eqref{chargesuper1} at this step, because $\hat L_{1}^{(\epsilon)}\frac{df}{d\omega}\neq 2\omega f$) and \eqref{energysuper1}.} In the limit $\epsilon\to 0$ we still get
\begin{equation}
Q_{p}^{(i)}[\xi_{i},\eta_{i},\gamma]+Q_{p}^{(j)}[\xi_{j},\eta_{j},\gamma],\qquad E_{p}^{(i)}[\xi_{i},\eta_{i},\gamma]+E_{p}^{(j)}[\xi_{j},\eta_{j},\gamma]
\end{equation}
without any cross terms. The latter implies that the additivity of the charge and the energy remains valid for the case  $\gamma_{i}=\gamma_{j}=\gamma$ with $\xi_{i}\not\equiv c\xi_{j}$, $\eta_{i}\not\equiv c\eta_{j}$, where $c$ is a constant, too.

Note that the symmetry between the modes with positive and negative charges, i.e., the symmetry $Q_{p}^{(n)}=-Q_{p}^{(k)}$, $E_{p}^{(n)}=E_{p}^{(k)}$, is not expected. An explicit example will be presented in Section~\ref{secexpexamp}.

One can see that if we drop the terms $\varrho_{\pm,nk}$ and $\theta_{\pm,nk}$ (which come as nonlinear corrections) in \eqref{chargeprelim}, we will get the nonzero time-dependent terms proportional to the overlap integrals between solutions for different modes. Thus, the use of the nonlinear equation of motion for perturbations not only recovers the conservation over time of the charge and the energy but also results in the additivity of the charge and the energy of the modes.

At the end of this section, let us return to the discussion of Q-ball stability in connection with the existence of the modes with $E_{p}^{(n)}[\xi_{n},\eta_{n},\gamma_{n}]<0$, which was started at the end of Section~\ref{subsoscmodes2}. As was shown in Section~\ref{subsdecmodes2}, the instability mode has $Q_{p}=0$, $E_{p}=0$; see \eqref{decmodechen}. So, there arises the question about a possibility to construct a perturbation with $Q_{p}=0$ and $E_{p}<0$, using the modes with $E_{p}^{(n)}[\xi_{n},\eta_{n},\gamma_{n}]<0$. If such perturbations exist, then the energy of the excited Q-ball will be less than the energy of the initial Q-ball. The latter may imply a potential instability of the Q-ball.

To this end, let us suppose that we have the modes with $E_{p}^{(n)}[\xi_{n},\eta_{n},\gamma_{n}]<0$, as well as the modes with $E_{p}^{(n)}[\xi_{n},\eta_{n},\gamma_{n}]>0$. Let us consider the perturbation constructed from these modes. According to \eqref{chargesuperpos}, for the total charge of the perturbation we have
\begin{equation}\label{totcharge0}
Q_{p}=\sum_{n}Q_{p}^{(n)}=0,
\end{equation}
For the energy, using \eqref{energysuperpos} and \eqref{negEpeq2}, we get
\begin{align}\nonumber
E_{p}=\sum_{n}E_{p}^{(n)}=\sum_{n}\left(
\omega Q_{p}^{(n)}+\alpha^{2}\int\Bigl(\gamma_{n}^{2}\xi_{n}^{*}\xi_{n}+\eta_{n}^{*}\hat L_{2}\eta_{n}\Bigr)\right)d^{d}x\\\label{enzerocharge}
=\omega\sum_{n}Q_{p}^{(n)}+
\alpha^{2}\sum_{n}\int\Bigl(\gamma_{n}^{2}\xi_{n}^{*}\xi_{n}+\eta_{n}^{*}\hat L_{2}\eta_{n}\Bigr)d^{d}x=
\alpha^{2}\sum_{n}\int\Bigl(\gamma_{n}^{2}\xi_{n}^{*}\xi_{n}+\eta_{n}^{*}\hat L_{2}\eta_{n}\Bigr)d^{d}x>0,
\end{align}
where we have used \eqref{totcharge0} and \eqref{negEpeq3}. We see that it is impossible to compose from the oscillation modes a perturbation with the charge $Q_{p}=0$ and $E_{p}<0$. This is an additional indication of the fact that the modes with $E_{p}^{(n)}[\xi_{n},\eta_{n},\gamma_{n}]<0$ do not induce new instabilities, on both classical and quantum levels. An explicit example of the mode with $E_{p}^{(n)}[\xi_{n},\eta_{n},\gamma_{n}]<0$ in the case of a classically stable Q-ball will be presented in Section~\ref{secexpexamp}.

Note that the last term of \eqref{enzerocharge} comes only from the second term of \eqref{energypertreduced}. This term does not contain any pathologies like time-dependent terms or the absence of additivity when only the linear part of the perturbation is taken into account (in the second term of \eqref{energypertreduced}, the corrections $\sim\alpha^{2}$ to the linear part of the perturbation lead to the terms $\sim\alpha^{3}$, which are neglected).

Analogously, using \eqref{dEdQ}, it is possible to show that the energy of any classically stable Q-ball of charge $Q_{0}+Q_{p}$ is smaller than the energy of the excited Q-ball of the same total charge $Q_{0}+Q_{p}$ (i.e., with the original charge $Q_{0}$ and the charge of the perturbation $Q_{p}$), regardless of the combination of the modes forming this perturbation:
\begin{align}\nonumber
\left(E_{0}[Q_{0}]+E_{p}\right)-E_{0}[Q_{0}+Q_{p}]\approx \left(E_{0}[Q_{0}]+E_{p}\right)-\left(E_{0}[Q_{0}]+\omega Q_{p}\right)\\
=E_{p}-\omega Q_{p}=\alpha^{2}\sum_{n}\int\Bigl(\gamma_{n}^{2}\xi_{n}^{*}\xi_{n}+\eta_{n}^{*}\hat L_{2}\eta_{n}\Bigr)d^{d}x>0.
\end{align}
This result also implies that the term ``excited Q-ball'' can be considered from a different point of view. Namely, if we somehow excite a Q-ball without changing its charge, i.e., if we just add only the energy to the system, it may look like a process changing the original Q-ball plus creating a perturbation,
\begin{equation}
E_{0}[Q_{0}]\to E_{0}[Q_{0}-Q_{p}]+E_{p}.
\end{equation}
It is clear that the change of the frequency of the Q-ball $\Delta\omega\sim\alpha^{2}$, and we can neglect the corrections $\Delta Q_{p}\sim\alpha^{4}$ and $\Delta E_{p}\sim\alpha^{4}$. The energy, which should be added to the initial Q-ball, is
\begin{align}\nonumber
\delta E=E_{p}-\omega Q_{p}=\alpha^{2}\sum_{n}\int\Bigl(\gamma_{n}^{2}\xi_{n}^{*}\xi_{n}+\eta_{n}^{*}\hat L_{2}\eta_{n}\Bigr)d^{d}x\\=\alpha^{2}\sum_{n}\int \Bigl(\gamma_{n}^{2}(\xi_{n}^{*}\xi_{n}+\eta_{n}^{*}\eta_{n})+\omega\gamma_{n}(\xi_{n}^{*}\eta_{n}+\eta_{n}^{*}\xi_{n})\Bigr)d^{d}x>0.
\end{align}
One can see that it is always positive. Again, $\delta E$ can be calculated using only the linear part of the perturbation.

\section{Isolation of a single mode}
For completeness, here I present a method of isolation of a single mode from the linear part of the perturbation, i.e., from superposition \eqref{pertlin}, which is a solution to linearized equations of motion \eqref{eqxin} and \eqref{eqetan}. Suppose we have a set of properly normalized solutions of Eqs.~\eqref{eqxin} and \eqref{eqetan}, which we define as $\hat\xi_{n}$ and $\hat\eta_{n}$. It is clear that since Eqs.~\eqref{eqxin} and \eqref{eqetan} are invariant with respect to the rescaling $\xi_{n}\to C_{n}\xi_{n}$ and $\eta_{n}\to C_{n}\eta_{n}$, the initial functions $a_{n}$ and $b_{n}$ in \eqref{pertlin} turn out to be rescaled as $a_{n}\to C_{n}a_{n}$ and $b_{n}\to C_{n}^{*}b_{n}$. Thus, the linear part of perturbation \eqref{pertlin} can be represented as
\begin{equation}\label{linpertnorm}
\varphi_{lin}(t,\vec x)=\sum_{n}\left(C_{n}\frac{1}{2}(\hat\xi_{n}+\hat\eta_{n})e^{i\gamma_{n} t}+C_{n}^{*}\frac{1}{2}(\hat\xi_{n}^{*}-\hat\eta_{n}^{*})e^{-i\gamma_{n} t}\right).
\end{equation}
The coefficients $C_{n}$ in \eqref{linpertnorm} can be obtained with the help of the formula
\begin{equation}\label{orthmodesQball}
C_{n}=
\frac{\int\left[(\hat\xi_{n}^{*}+\hat\eta_{n}^{*})\left((\gamma_{n}+2\omega)\varphi_{lin}-i\dot\varphi_{lin}\right)+
(\hat\xi_{n}^{*}-\hat\eta_{n}^{*})\left((\gamma_{n}-2\omega)\varphi_{lin}^{*}-i\dot\varphi_{lin}^{*}\right)\right]d^{d}x}{2e^{i\gamma_{n} t}
\int\left[\gamma_{n}(\hat\xi_{n}^{*}\hat\xi_{n}+\hat\eta_{n}^{*}\hat\eta_{n})+\omega(\hat\xi_{n}^{*}\hat\eta_{n}+\hat\xi_{n}\hat\eta_{n}^{*})\right]d^{d}x},
\end{equation}
which can be evaluated at any moment of time $t$. The derivation of Eq.~\eqref{orthmodesQball} is straightforward and relies on the use of the complex conjugate of Eqs.~\eqref{appCeq1} and \eqref{appCeq2}. In fact, Eqs.~\eqref{appCeq1} and \eqref{appCeq2} exactly suggest the form of transform \eqref{orthmodesQball}.

\section{Explicit example: logarithmic scalar field potential}\label{secexpexamp}
Let us consider $d=3$ and the scalar field potential of the form \cite{Rosen1}
\begin{equation}\label{logscpot}
V(\phi^{*}\phi)=-\mu^{2}\phi^{*}\phi\ln\left(\beta^{2}\phi^{*}\phi\right).
\end{equation}
The Q-ball profile in this model has the form
\begin{equation}\label{logqballsol}
f(r)=\frac{1}{\beta}e^{-\frac{\omega^{2}}{2\mu^{2}}+1}e^{-\frac{\mu^{2}r^{2}}{2}}.
\end{equation}
The charge and the energy of this Q-ball look like
\begin{equation}
Q_{0}=2\pi^{\frac{3}{2}}\frac{\omega}{\beta^{2}\mu^{3}}\,\textrm{e}^{2-\frac{\omega^2}{\mu^2}},\qquad
E_{0}=2\pi^{\frac{3}{2}}\frac{1}{\beta^{2}\mu}\left(\frac{\omega^{2}}{\mu^{2}}+\frac{1}{2}\right)\textrm{e}^{2-\frac{\omega^2}{\mu^2}}.
\end{equation}

The model with potential \eqref{logscpot} is known to be exactly solvable --- not only Q-ball solution \eqref{logqballsol} can be found analytically \cite{Rosen1} but also the linearized equations of motion can be solved analytically \cite{MarcVent}, providing detailed information about the spectrum of perturbations. In particular, it was shown in paper \cite{MarcVent} that Q-balls in this model are classically stable for $\omega^{2}\ge\frac{\mu^{2}}{2}$.

For \eqref{UGdef}, \eqref{Jdef}, and $\frac{df}{d\omega}$ we get
\begin{equation}\label{logpotentparam}
U=-3\mu^{2}+\omega^{2}+\mu^{4}r^{2},\qquad S=-\mu^{2},\qquad J=\frac{\mu^{2}}{2f},\qquad \frac{df}{d\omega}=-\frac{\omega}{\mu^{2}}f.
\end{equation}
As was noted above, the linearized equations of motion for this model were solved analytically in \cite{MarcVent}. Here I will repeat only the main steps of calculations in order to get the formulas in our notations. Equations \eqref{eqxin} and \eqref{eqetan} take the form
\begin{align}\label{loglineq1}
-\Delta\xi_{n}+\mu^{4}r^{2}\xi_{n}=\gamma_{n}^{2}\xi_{n}+5\mu^{2}\xi_{n}+2\omega\gamma_{n}\eta_{n},\\\label{loglineq2}
-\Delta\eta_{n}+\mu^{4}r^{2}\eta_{n}=\gamma_{n}^{2}\eta_{n}+3\mu^{2}\eta_{n}+2\omega\gamma_{n}\xi_{n}.
\end{align}
As was noted in Section~\ref{pertsection}, Eqs.~\eqref{loglineq1} and \eqref{loglineq2} can be used for instability modes too if $\xi_{n}$ is supposed to be real, whereas $\eta_{n}$ is supposed to be imaginary. It is clear that solutions to these equations are connected with the eigenfunctions of the operator $-\Delta+\mu^{4}r^{2}$, which is just the Hamiltonian of the quantum mechanical harmonic oscillator,
\begin{equation}
-\Delta\Psi_{\hat N}(\vec x)+\mu^{4}r^{2}\Psi_{\hat N}(\vec x)=\mu^{2}(3+2N)\Psi_{\hat N}(\vec x),
\end{equation}
where $\hat N=\{N_{1}, N_{2}, N_{3}\}$, $N=N_{1}+N_{2}+N_{3}$, and $N_{1,2,3}=0,1,2,...$. So, we define
\begin{equation}
\xi_{\hat N,\pm}(\vec x)=\Psi_{\hat N}(\vec x)Y_{\hat N,\pm},\qquad \eta_{\hat N,\pm}(\vec x)=\Psi_{\hat N}(\vec x)Z_{\hat N,\pm},
\end{equation}
where the subscripts $\hat N,\pm$ are used instead of the subscript $n$ in \eqref{loglineq1} and \eqref{loglineq2}. The meaning of the subscripts ``+'' and ``--'' will become clear later. Here $Y_{\hat N,\pm}$ and $Z_{\hat N,\pm}$ are complex coefficients, for which we get
\begin{align}
(\gamma_{N,\pm}^{2}+2\mu^{2}-2N\mu^{2})Y_{\hat N,\pm}+2\omega\gamma_{N,\pm}Z_{\hat N,\pm}=0,\\\label{logyzeq}
2\omega\gamma_{N,\pm}Y_{\hat N,\pm}+(\gamma_{N,\pm}^{2}-2N\mu^{2})Z_{\hat N,\pm}=0,
\end{align}
leading to
\begin{equation}\label{gammasol}
\gamma_{N,\pm}^{2}=2\mu^{2}\left(\frac{\omega^{2}}{\mu^{2}}-\frac{1}{2}+N\pm\sqrt{\left(\frac{\omega^{2}}{\mu^{2}}-\frac{1}{2}\right)^{2}+2N\frac{\omega^{2}}{\mu^{2}}}\right).
\end{equation}
Note that there is the subscript $N$ in $\gamma_{N,\pm}$, not $\hat N$.

According to \eqref{gammasol}, for $\omega^{2}>\frac{\mu^{2}}{2}$ there are two solutions corresponding to the modes with the zero frequency (time-independent modes):
\begin{equation}
\gamma_{0,-}=0,\qquad \gamma_{1,-}=0
\end{equation}
(in order to correctly use the notations $Y_{\hat N,\pm}$ and $Z_{\hat N,\pm}$ for the modes with $\gamma_{N,\pm}=0$, it is also necessary to restrict $Y_{\hat N,\pm}$ to be purely real and $Z_{\hat N,\pm}$ to be purely imaginary for these modes, as in the case of instability modes). The first solution corresponds to the mode $\alpha if$, whereas the second solution corresponds to the translational modes $\alpha\partial_{i}f$, $i=1,2,3$. It is not difficult to show that $\gamma_{N,\pm}^{2}>0$ for the other modes if $\omega^{2}>\frac{\mu^{2}}{2}$. Without loss of generality, we can suppose that $\gamma_{N,\pm}>0$ for these modes. It is clear that the modes with the same $\gamma_{N,+}$, as well as the modes with the same $\gamma_{N,-}$, are degenerate --- they have the same value of the frequency but different ``wave functions'' $\Psi_{\hat N}$.

For $\omega^{2}<\frac{\mu^{2}}{2}$ the $(0,-)$ mode turns out to be the instability mode with
\begin{equation}
\gamma_{0,-}^{2}=2(2\omega^{2}-\mu^{2})<0.
\end{equation}

Now we turn to the charge and the energy. Using \eqref{logpotentparam}, from \eqref{chargesuper1} and \eqref{energysuper1} we get for each mode with $\gamma_{N,\pm}>0$
\begin{equation}\label{chargelogprelim}
Q_{p}^{\hat N,\pm}=\alpha^{2}\int \gamma_{N,\pm}\left(\xi_{\hat N,\pm}\eta_{\hat N,\pm}^{*}+\eta_{\hat N,\pm}\xi_{\hat N,\pm}^{*}\right)d^{3}x,
\end{equation}
\begin{equation}\label{energylogprelim}
E_{p}^{\hat N,\pm}=2\omega Q_{p}^{\hat N,\pm}+\alpha^{2}\int
\gamma_{N,\pm}^{2}\left(\xi_{\hat N,\pm}\xi_{\hat N,\pm}^{*}+\eta_{\hat N,\pm}\eta_{\hat N,\pm}^{*}\right)d^{3}x.
\end{equation}
From, say, Eq.~\eqref{logyzeq}, we get for $\omega\neq 0$
\begin{equation}
Y_{\hat N,\pm}=-\frac{\gamma_{N,\pm}^{2}-2N\mu^{2}}{2\omega\gamma_{N,\pm}}Z_{\hat N,\pm}.
\end{equation}
Substituting the latter relation into \eqref{chargelogprelim} and \eqref{energylogprelim}, we obtain
\begin{align}\label{chargelog}
Q_{p}^{\hat N,\pm}=-\alpha^{2}\frac{2\mu^{2}}{\omega}\left(\frac{\omega^{2}}{\mu^{2}}-\frac{1}{2}\pm\sqrt{\left(\frac{\omega^{2}}{\mu^{2}}-
\frac{1}{2}\right)^{2}+2N\frac{\omega^{2}}{\mu^{2}}}\right)|Z_{\hat N,\pm}|^{2}\int|\Psi_{\hat N}|^{2}d^{3}x,\\
\label{energylog}
E_{p}^{\hat N,\pm}=\alpha^{2}\frac{\mu^{4}}{\omega^{2}}
\left(4N\frac{\omega^{2}}{\mu^{2}}-\left(\frac{\omega^{2}}{\mu^{2}}-\frac{1}{2}\right)\mp\sqrt{\left(\frac{\omega^{2}}{\mu^{2}}-
\frac{1}{2}\right)^{2}+2N\frac{\omega^{2}}{\mu^{2}}}\right)|Z_{\hat N,\pm}|^{2}\int|\Psi_{\hat N}|^{2}d^{3}x.
\end{align}
Since $\int d^{3}x\,\Psi_{\hat N}^{*}\Psi_{\hat K}=0$ for $\hat N\neq\hat K$, there cannot be cross terms in the expressions for the charge and the energy of the perturbation containing the modes of the same frequency but with different wave functions. The latter supports the conclusion of Section~6, which is related to Eqs.~\eqref{supeqfreq1} and \eqref{supeqfreq2}.

We see from \eqref{chargelog} that the sign of $Q_{p}^{\hat N,\pm}$ can be positive or negative, depending on the sign $+$ or $-$ in \eqref{chargelog} and the sign of $\omega$. In general, there is no symmetry $Q_{p}^{(n)}=-Q_{p}^{(k)}$, $E_{p}^{(n)}=E_{p}^{(k)}$. The easiest way to see it is to consider $\omega=\frac{\mu}{\sqrt{2}}>0$. In this case, for the nonzero $\gamma_{N,\pm}$ we can write
\begin{align}
&Q_{p}^{\hat N,\pm}=\mp\sqrt{N}C_{\hat N,\pm},\\
&E_{p}^{\hat N,\pm}=\frac{\mu}{\sqrt{2}}\left(2N\mp\sqrt{N}\right)C_{\hat N,\pm},
\end{align}
where $C_{\hat N,\pm}>0$. Requirements for the symmetry under consideration look as follows:
\begin{align}
\sqrt{N}C_{\hat N,+}&=\sqrt{K}C_{\hat K,-},\\
\left(2N-\sqrt{N}\right)C_{\hat N,+}&=\left(2K+\sqrt{K}\right)C_{\hat K,-},
\end{align}
resulting in
\begin{equation}
\sqrt{N}=1+\sqrt{K}.
\end{equation}
Of course, this equation has solutions for some natural $N$ and $K$, but obviously not for all.

It is interesting to note that for $\omega^{2}>\frac{\mu^{2}}{2}$, $N=0$, and the sign $-$ in \eqref{energylog} we get
\begin{equation}\label{enoscmodelog}
E_{p}^{0,+}=-2\alpha^{2}\frac{\mu^{4}}{\omega^{2}}\left(\frac{\omega^{2}}{\mu^{2}}-\frac{1}{2}\right)
|Z_{0,+}|^{2}\int|\Psi_{0}|^{2}\,d^{3}x<0,
\end{equation}
where the subscript $\hat N=\{0,0,0\}$ is replaced by ``0'', which corresponds to the oscillation mode with
\begin{equation}
\gamma_{0,+}=\sqrt{2}\sqrt{2\omega^{2}-\mu^{2}}.
\end{equation}
We see that the energy of the oscillation mode indeed can be negative even in the case of a classically stable Q-ball. For this mode,
\begin{equation}
|E_{p}^{0,+}|=\frac{\mu^{2}}{2\omega^{2}}|\omega Q_{p}^{0,+}|<|\omega Q_{p}^{0,+}|,
\end{equation}
which corresponds to \eqref{negEpeq4}. If we pass from the stable region $\omega^{2}>\frac{\mu^{2}}{2}$ to the unstable region $\omega^{2}<\frac{\mu^{2}}{2}$, this mode will transform (through the time-independent mode $\sim if$) into the instability mode with $\gamma_{0,-}=-i\sqrt{2}\sqrt{\mu^{2}-2\omega^{2}}$; see the discussion at the end of Section~\ref{pertsection}. Since the energy of an instability mode is equal to zero, the energy of the corresponding oscillation mode should tend to zero as $\gamma_{n}\to 0$. It is easy to see that indeed $E_{p}^{0,+}\to 0$ as $\gamma_{0,+}\to 0$. However, one cannot expect that in the general case the energy of the oscillation mode, which transforms into the instability mode, is always negative.

As for the other modes with $\gamma_{N,\pm}>0$ in this model, the inequality $E_{p}^{\hat N,\pm}>0$ holds for these modes. For $N\gg 1$ and $N\gg \frac{\omega^{2}}{\mu^{2}}$ we can get $E_{p}^{\hat N,\pm}\approx\mp\mu\sqrt{2}\sqrt{N}\,Q_{p}^{\hat N,\pm}\sim N$.

\section{Conclusion}
In the present paper, the charge and the energy of perturbations against a Q-ball solution are considered up to the second order in perturbations. It is shown that in order to obtain correct results (in particular, to maintain the conservation over time of the charge and the energy), it is necessary to consider nonlinear equations of motion for perturbations, i.e., equations which include quadratic terms in perturbations. It is also shown that, contrary to the naive expectation for the nonlinear theory, the additivity property is valid for the charge and the energy if the perturbation against a Q-ball contains many modes. It is expected that the same is true for other integrals of motion such as momentum or angular momentum.

This additivity of the charge and the energy can be explained in the following way. Indeed, the overlap integrals for different modes in $Q_{p}$ and $E_{p}$, at least for the modes with different frequencies $\gamma_{n}$, can appear together only with the oscillating exponents $e^{\pm i(\gamma_{n}\pm\gamma_{k})t}$. Since the charge and the energy are conserved over time up to the second order in perturbations (due to the use of the nonlinear equation of motion for the perturbation), such terms should vanish.

In fact, the analysis presented in this paper is just the correct use of the perturbation theory. However, apart from the main results, it reveals some interesting consequences:
\begin{enumerate}
\item
If one knows not only the Q-ball profile $f(r)$ but also $\frac{df(r)}{d\omega}$, there is no need to solve the equations of motion for the nonlinear correction: the charge and the energy of the mode can be obtained using only the solutions to linearized equations of motion. This result can be useful for consistent and correct quantization of perturbations against Q-balls.
\item
The energy of the oscillation mode can be negative (with respect to the Q-ball energy), which was demonstrated explicitly for the model with logarithmic scalar field potential. However, it does not lead to additional instabilities of the Q-ball with such modes in the spectrum of perturbations.
\item
The second order corrections to the linear part of a perturbation are not necessary if we consider the perturbation that does not change the total charge of the system (i.e., with $Q_{p}=0$). In such a case, the energy of the perturbation (in the quadratic order in the expansion parameter $\alpha$) can be calculated using only the linear part of the perturbation, which does not lead to pathologies such as emergence of time-dependent terms or violation of the additivity property.
\item
Nonlinear corrections are not necessary for Q-balls with $\omega=0$, i.e., for static background solutions. Meanwhile, it is necessary to note that all Q-balls with $\omega=0$ are classically unstable (see, for example, \cite{Panin:2016ooo}).
\item
The form of oscillation mode \eqref{substosccorr} with nonlinear corrections implies that in the full nonlinear theory there may exist exact solutions for small perturbations $\varphi$, representing the nonlinear modes that are periodic in time with the periods $T_{n}=\frac{2\pi}{\gamma_{n}}$ (here I do not take into account the overall factor $e^{i\omega t}$). Although it is not obvious that the full nonlinear theory indeed admits the existence of such solutions, if they still exist, the corresponding terms of their Fourier transform will give the result close to Eq.~\eqref{substosccorr}.
\end{enumerate}

In principle, an analogous situation with the necessity to consider nonlinear equations of motion for perturbations may appear in other theories with time-dependent background solutions (for the case of the nonlinear Schr\"{o}dinger equation, see \cite{Smolyakov}). Thus, I hope that the approach and the results presented in this paper can be useful not only for examining Q-balls but also for studying perturbations against other time-dependent background solutions.

\section*{Acknowledgements}
The author is grateful to D.G.~Levkov, E.Ya.~Nugaev and I.P.~Volobuev for valuable discussions. The work was supported by Grant No. 16-12-10494 of the Russian Science Foundation.

\section*{Appendix~A: Perturbations in the case of small $|\rho|$}
Let us substitute ansatz \eqref{pertcomplex1} and \eqref{pertcomplex2} into linearized equations of motion \eqref{lineq1intro} and \eqref{lineq2intro}. Using the relations $\hat L_{2}f=0$ and Eq.~\eqref{dfdomegaL1}, we get
\begin{align}\label{App0eq1}
&\hat L_{1}c_{1}-2\omega c_{2}-\frac{df}{d\omega}-\rho^{2}c_{1}=0,\\ \label{App0eq2}
&\hat L_{2}c_{2}-2\omega \frac{df}{d\omega}-f-\rho^{2}(2\omega c_{1}+c_{2})=0.
\end{align}
We see that $c_{1}(\vec x)$ and $c_{2}(\vec x)$ are not singular in $\rho^{2}$. Now let us make the following steps. First, let us multiply Eq.~\eqref{App0eq2} by $f$ and integrate the result over the space. We get
\begin{equation}\label{App0eq3}
\frac{1}{2}\frac{dQ_{0}}{d\omega}+\rho^{2}\int(2\omega c_{1}f+c_{2}f)d^{d}x=0,
\end{equation}
where we have used the definition of Q-ball charge \eqref{qballcharge}. Second, let us take the complex conjugate of Eq.~\eqref{App0eq2}, multiply it by $c_{2}(\vec x)$, and integrate the result over the space. We get
\begin{equation}\label{App0eq4}
\int c_{2}\hat L_{2}c_{2}^{*}\,d^{d}x=\int\left(2\omega c_{2}\frac{df}{d\omega}+c_{2}f\right)d^{d}x+{\rho^{*}}^{2}\int\left(2\omega c_{1}^{*}c_{2}f+c_{2}^{*}c_{2}f\right)d^{d}x.
\end{equation}
And third, let us multiply Eq.~\eqref{App0eq1} by $\frac{df}{d\omega}$ and integrate the result over the space. Using Eq.~\eqref{dfdomegaL1}, we get
\begin{equation}\label{App0eq5}
\int\left(\frac{df}{d\omega}\right)^{2}d^{d}x=\int\left(2\omega c_{1}f-2\omega c_{2}\frac{df}{d\omega}\right)d^{d}x-\rho^{2}\int c_{1}\frac{df}{d\omega}d^{d}x.
\end{equation}
After summing up \eqref{App0eq4} and \eqref{App0eq5}, substituting the result into \eqref{App0eq3}, and omitting the terms $\sim\rho^{4}$ and $\sim{\rho^{*}}^{2}\rho^{2}$, we get relation \eqref{stinstsmallrho}.

\section*{Appendix~B: The integrals arising when calculating the charge, single mode}
Let us take Eq.~\eqref{singpartpsi2}, multiply it by $f$, and integrate the result over the space. Using the fact that $\hat L_{2}f=0$, we get
\begin{equation}\label{appA1}
\int(4\omega f\psi_{+}+4\gamma f\psi_{-})d^{d}x=\frac{1}{\gamma}\int S\xi\eta\, d^{d}x.
\end{equation}
Then, using Eqs.~\eqref{lineq1} and \eqref{lineq2}, let us consider the integrals
\begin{align}
\int \eta(\hat L_{1}\xi)d^{d}x=\int \eta(2\omega\gamma\eta+\gamma^{2}\xi)d^{d}x,\\
\int \xi(\hat L_{2}\eta)d^{d}x=\int \xi(2\omega\gamma\xi+\gamma^{2}\eta)d^{d}x.
\end{align}
Since from definition of the operators $\hat L_{1}$ and $\hat L_{2}$, \eqref{L1eq} and \eqref{L2eq}, it follows that
\begin{equation}\label{appBLdefeq}
\int(\eta\hat L_{1}\xi-\xi\hat L_{2}\eta)d^{d}x=2\int S\xi\eta\, d^{d}x,
\end{equation}
we get
\begin{equation}\label{appA2}
\frac{1}{\gamma}\int S\xi\eta\, d^{d}x=\omega\int(\eta^{2}-\xi^{2})d^{d}x.
\end{equation}
Combining \eqref{appA1} and \eqref{appA2}, we obtain
\begin{equation}
\int\Bigl(4\omega f\psi_{+}+4\gamma f\psi_{-}+\omega(\xi^{2}-\eta^{2})\Bigr)d^{d}x=0,
\end{equation}
which is exactly relation \eqref{appAeq}.

\section*{Appendix~C: The integrals arising when calculating the charge, many modes}
Let us take Eq.~\eqref{eqvarrho}, multiply it by $f$, and integrate the result over the space. Using the fact that $\hat L_{2}f=0$, we get
\begin{equation}\label{auxil0}
(\gamma_{n}+\gamma_{k})\int\left(2\omega f\varrho_{+,nk}+(\gamma_{n}+\gamma_{k})f\varrho_{-,nk}\right)d^{d}x=
\int S\left(\xi_{n}\eta_{k}+\eta_{n}\xi_{k}\right)d^{d}x.
\end{equation}
Then let us take Eqs.~\eqref{eqxin} and \eqref{eqetan} and consider the integrals
\begin{align}
\int \eta_{k}(\hat L_{1}\xi_{n})d^{d}x=\int \eta_{k}(2\omega\gamma_{n}\eta_{n}+\gamma_{n}^{2}\xi_{n})d^{d}x,\\
\int \xi_{n}(\hat L_{2}\eta_{k})d^{d}x=\int \xi_{n}(2\omega\gamma_{k}\xi_{k}+\gamma_{k}^{2}\eta_{k})d^{d}x.
\end{align}
Since
\begin{equation}
\int (\eta_{k}\hat L_{1}\xi_{n}-\xi_{n}\hat L_{2}\eta_{k})d^{d}x=2\int S\xi_{n}\eta_{k}\,d^{d}x,
\end{equation}
we get
\begin{equation}\label{auxil1}
2\int S\xi_{n}\eta_{k}\,d^{d}x=\int\left(2\omega(\gamma_{n}\eta_{n}\eta_{k}-\gamma_{k}\xi_{n}\xi_{k})+(\gamma_{n}^{2}-\gamma_{k}^{2})\xi_{n}\eta_{k}\right)d^{d}x.
\end{equation}
Now let us take Eqs.~\eqref{eqxin} and \eqref{eqetan} and consider the integrals
\begin{align}
\int\eta_{n}(\hat L_{1}\xi_{k})d^{d}x=\int\eta_{n}(2\omega\gamma_{k}\eta_{k}+\gamma_{k}^{2}\xi_{k})d^{d}x,\\
\int\xi_{k}(\hat L_{2}\eta_{n})d^{d}x=\int \xi_{k}(2\omega\gamma_{n}\xi_{n}+\gamma_{n}^{2}\eta_{n})d^{d}x.
\end{align}
Analogously, we get
\begin{equation}\label{auxil2}
2\int S\xi_{k}\eta_{n}\,d^{d}x=\int\left(2\omega(\gamma_{k}\eta_{n}\eta_{k}-\gamma_{n}\xi_{n}\xi_{k})+(\gamma_{k}^{2}-\gamma_{n}^{2})\xi_{k}\eta_{n}\right)d^{d}x.
\end{equation}
Summing up \eqref{auxil1} and \eqref{auxil2}, we obtain
\begin{equation}\label{auxil3}
\int S(\xi_{n}\eta_{k}+\xi_{k}\eta_{n})d^{d}x=\int \left(\omega(\gamma_{n}+\gamma_{k})(\eta_{n}\eta_{k}-\xi_{n}\xi_{k})+\frac{1}{2}(\gamma_{n}^{2}-\gamma_{k}^{2})(\xi_{n}\eta_{k}-\xi_{k}\eta_{n})\right)d^{d}x.
\end{equation}
Finally, combining \eqref{auxil0} and \eqref{auxil3} and using the fact that $\gamma_{n}+\gamma_{k}\neq 0$, we arrive at
\begin{equation}
\int \left(2\omega f\varrho_{+,nk}+(\gamma_{n}+\gamma_{k})f\varrho_{-,nk}+
\omega(\xi_{n}\xi_{k}-\eta_{n}\eta_{k})+\frac{1}{2}(\gamma_{n}-\gamma_{k})(\xi_{k}\eta_{n}-\xi_{n}\eta_{k})\right)d^{d}x=0,
\end{equation}
which is exactly relation \eqref{appBeq1}.

Relation \eqref{appBeq2} can be obtained in an analogous way. Multiplying Eq.~\eqref{eqtheta} by $f$ and integrating the result over the space, we get
\begin{equation}\label{auxil4}
(\gamma_{n}-\gamma_{k})\int \left(2\omega f\theta_{+,nk}+(\gamma_{n}-\gamma_{k})f\theta_{-,nk}\right)d^{d}x=\int S\left(\eta_{n}\xi_{k}^{*}-\xi_{n}\eta_{k}^{*}\right)d^{d}x.
\end{equation}
Then let us take Eqs.~\eqref{eqxin} and \eqref{eqetan} and consider the integrals
\begin{align}
\int \eta_{n}(\hat L_{1}\xi_{k}^{*})d^{d}x=\int \eta_{n}(2\omega\gamma_{k}\eta_{k}^{*}+\gamma_{k}^{2}\xi_{k}^{*})d^{d}x,\\
\int \xi_{k}^{*}(\hat L_{2}\eta_{n})d^{d}x=\int \xi_{k}^{*}(2\omega\gamma_{n}\xi_{n}+\gamma_{n}^{2}\eta_{n})d^{d}x,
\end{align}
leading to
\begin{equation}\label{auxil5}
2\int S\xi_{k}^{*}\eta_{n}\,d^{d}x=\int \left(2\omega(\gamma_{k}\eta_{n}\eta_{k}^{*}-\gamma_{n}\xi_{n}\xi_{k}^{*})+(\gamma_{k}^{2}-\gamma_{n}^{2})\xi_{k}^{*}\eta_{n}\right)d^{d}x,
\end{equation}
and the integrals
\begin{align}
\int \eta_{k}^{*}(\hat L_{1}\xi_{n})d^{d}x=\int \eta_{k}^{*}(2\omega\gamma_{n}\eta_{n}+\gamma_{n}^{2}\xi_{n})d^{d}x,\\
\int \xi_{n}(\hat L_{2}\eta_{k}^{*})d^{d}x=\int \xi_{n}(2\omega\gamma_{k}\xi_{k}^{*}+\gamma_{k}^{2}\eta_{k}^{*})d^{d}x,
\end{align}
leading to
\begin{equation}\label{auxil6}
2\int S\xi_{n}\eta_{k}^{*}\,d^{d}x=\int \left(2\omega(\gamma_{n}\eta_{n}\eta_{k}^{*}-\gamma_{k}\xi_{n}\xi_{k}^{*})+(\gamma_{n}^{2}-\gamma_{k}^{2})\xi_{n}\eta_{k}^{*}\right)d^{d}x.
\end{equation}
Subtracting \eqref{auxil6} from \eqref{auxil5}, we obtain
\begin{equation}\label{auxil9}
\int S(\xi_{k}^{*}\eta_{n}-\xi_{n}\eta_{k}^{*})d^{d}x=\int \left(\omega(\gamma_{k}-\gamma_{n})(\eta_{n}\eta_{k}^{*}+\xi_{n}\xi_{k}^{*})
+\frac{1}{2}(\gamma_{k}^{2}-\gamma_{n}^{2})(\xi_{k}^{*}\eta_{n}+\xi_{n}\eta_{k}^{*})\right)d^{d}x.
\end{equation}
Combining \eqref{auxil4} and \eqref{auxil9} and using the fact that $\gamma_{n}-\gamma_{k}\neq 0$, we arrive at
\begin{equation}
\int \left(2\omega f\theta_{+,nk}+(\gamma_{n}-\gamma_{k})f\theta_{-,nk}+\omega(\eta_{n}\eta_{k}^{*}+\xi_{n}\xi_{k}^{*})+
\frac{1}{2}(\gamma_{n}+\gamma_{k})(\xi_{k}^{*}\eta_{n}+\xi_{n}\eta_{k}^{*})\right)d^{d}x=0,
\end{equation}
which is exactly relation \eqref{appBeq2}.

\section*{Appendix~D: The integrals arising when calculating the energy, many modes}
First, using Eq.~\eqref{eqxin}, let us consider the integrals
\begin{align}
\int \xi_{k}(\hat L_{1}\xi_{n})d^{d}x=\int \xi_{k}(2\omega\gamma_{n}\eta_{n}+\gamma_{n}^{2}\xi_{n})d^{d}x,\\
\int \xi_{n}(\hat L_{1}\xi_{k})d^{d}x=\int \xi_{n}(2\omega\gamma_{k}\eta_{k}+\gamma_{k}^{2}\xi_{k})d^{d}x,
\end{align}
leading to
\begin{equation}\label{auxen1}
\int \left(2\omega\gamma_{n}\eta_{n}\xi_{k}-2\omega\gamma_{k}\eta_{k}\xi_{n}+(\gamma_{n}^{2}-\gamma_{k}^{2})\xi_{n}\xi_{k}\right)d^{d}x=0.
\end{equation}
Then, using Eq.~\eqref{eqetan}, let us consider the integrals
\begin{align}
\int \eta_{k}(\hat L_{2}\eta_{n})d^{d}x=\int \eta_{k}(2\omega\gamma_{n}\xi_{n}+\gamma_{n}^{2}\eta_{n})d^{d}x,\\
\int \eta_{n}(\hat L_{2}\eta_{k})d^{d}x=\int \eta_{n}(2\omega\gamma_{k}\xi_{k}+\gamma_{k}^{2}\eta_{k})d^{d}x,
\end{align}
leading to
\begin{equation}\label{auxen2}
\int \left(2\omega\gamma_{n}\eta_{k}\xi_{n}-2\omega\gamma_{k}\eta_{n}\xi_{k}+(\gamma_{n}^{2}-\gamma_{k}^{2})\eta_{n}\eta_{k}\right)d^{d}x=0.
\end{equation}
Combining \eqref{auxen1} and \eqref{auxen2}, we obtain
\begin{equation}
\int \left(2\omega(\gamma_{n}+\gamma_{k})(\eta_{n}\xi_{k}-\eta_{k}\xi_{n})+(\gamma_{n}^{2}-\gamma_{k}^{2})(\xi_{n}\xi_{k}-\eta_{n}\eta_{k})\right)d^{d}x=0.
\end{equation}
Since $\gamma_{n}+\gamma_{k}\neq 0$, we get
\begin{equation}
\int\left(2\omega(\eta_{n}\xi_{k}-\eta_{k}\xi_{n})+(\gamma_{n}-\gamma_{k})(\xi_{n}\xi_{k}-\eta_{n}\eta_{k})\right)d^{d}x=0,
\end{equation}
which is exactly relation \eqref{appCeq1}.

Relation \eqref{appCeq2} can be obtained in an analogous way. Using Eq.~\eqref{eqxin}, we consider the integrals
\begin{align}
\int \xi_{k}^{*}(\hat L_{1}\xi_{n})d^{d}x=\int \xi_{k}^{*}(2\omega\gamma_{n}\eta_{n}+\gamma_{n}^{2}\xi_{n})d^{d}x,\\
\int \xi_{n}(\hat L_{1}\xi_{k}^{*})d^{d}x=\int \xi_{n}(2\omega\gamma_{k}\eta_{k}^{*}+\gamma_{k}^{2}\xi_{k}^{*})d^{d}x,
\end{align}
leading to
\begin{equation}\label{auxen3}
\int \left(2\omega\gamma_{n}\eta_{n}\xi_{k}^{*}-2\omega\gamma_{k}\eta_{k}^{*}\xi_{n}+(\gamma_{n}^{2}-\gamma_{k}^{2})\xi_{n}\xi_{k}^{*}\right)d^{d}x=0.
\end{equation}
Then, using Eq.~\eqref{eqetan}, we take the integrals
\begin{align}
\int \eta_{k}^{*}(\hat L_{2}\eta_{n})d^{d}x=\int \eta_{k}^{*}(2\omega\gamma_{n}\xi_{n}+\gamma_{n}^{2}\eta_{n})d^{d}x,\\
\int \eta_{n}(\hat L_{2}\eta_{k}^{*})d^{d}x=\int \eta_{n}(2\omega\gamma_{k}\xi_{k}^{*}+\gamma_{k}^{2}\eta_{k}^{*})d^{d}x,
\end{align}
leading to
\begin{equation}\label{auxen4}
\int \left(2\omega\gamma_{n}\eta_{k}^{*}\xi_{n}-2\omega\gamma_{k}\eta_{n}\xi_{k}^{*}+(\gamma_{n}^{2}-\gamma_{k}^{2})\eta_{n}\eta_{k}^{*}\right)d^{d}x=0.
\end{equation}
Combining \eqref{auxen3} and \eqref{auxen4}, we obtain
\begin{equation}
\int \left(2\omega(\gamma_{n}-\gamma_{k})(\eta_{n}\xi_{k}^{*}+\xi_{n}\eta_{k}^{*})+(\gamma_{n}^{2}-\gamma_{k}^{2})(\xi_{n}\xi_{k}^{*}+\eta_{n}\eta_{k}^{*})\right)d^{d}x=0.
\end{equation}
Since $\gamma_{n}-\gamma_{k}\neq 0$, we get
\begin{equation}
\int \left(2\omega(\eta_{n}\xi_{k}^{*}+\xi_{n}\eta_{k}^{*})+(\gamma_{n}+\gamma_{k})(\xi_{n}\xi_{k}^{*}+\eta_{n}\eta_{k}^{*})\right)d^{d}x=0,
\end{equation}
which is exactly relation \eqref{appCeq2}.

\end{document}